%% file: Wignerv2.tex
\documentclass[11pt,a4paper]{article}
\usepackage{mathptmx}
\usepackage{amsmath, amssymb,amscd,mathtools,braket,mathrsfs,commath}
\usepackage{enumitem}
\usepackage{tikz-cd}
\usepackage{helvet}
\usepackage{courier}
\usepackage{float}
\usepackage{type1cm}

\input{Bohrificationpreamble.tex}

\usepackage{stmaryrd}
\newcommand{\J}{\mathsf{J}}
\renewcommand{\K}{\mathsf{K}}
\newcommand{\W}{\mathsf{W}}
\newtheorem{definition}{Definition}[section]
\newtheorem{lemma}[definition]{Lemma}
\newtheorem{theorem}[definition]{Theorem}
\newtheorem{proposition}[definition]{Proposition}

\newtheorem{corollary}[definition]{Corollary}
\numberwithin{equation}{section}
\topmargin = - 1 cm \textheight = 23 cm \textwidth = 15 cm
\begin{document}
\pagenumbering{arabic} \setlength{\unitlength}{1cm}\cleardoublepage
\date\nodate
\begin{center}
\begin{huge}
{\bf (No) Wigner Theorem for C*-algebras}
\end{huge}
\bigskip

\renewcommand{\thefootnote}{\alph{footnote}}
\begin{Large}
 Klaas Landsman\footnote{ 
Institute for Mathematics, Astrophysics, and Particle Physics (IMAPP), Radboud University, Nijmegen, and \\  Dutch Institute for Emergent Phenomena (DIEP), \texttt{www.d-iep.org}. Email: \texttt{landsman@math.ru.nl}.} 
\hspace{10pt} and \hspace{10pt} Kitty Rang\footnote{De Nederlandsche Bank (DNB), Westeinde 1, 1017 ZN Amsterdam. Email:
\texttt{kitty.r.rang@gmail.com}. }
\end{Large}
\bigskip

\end{center}
\bigskip

 \begin{abstract} 
\noindent 
Wigner's Theorem states that bijections of the set $\CP_1(H)$ of one-dimensional projections on a
 \Hs\  $H$ that preserve transition probabilities are induced by either a unitary or an anti-unitary operator 
on $H$ (which is uniquely determined up to a phase). Since elements of $\CP_1(H)$ define pure states on the C*-algebra $B(H)$ of all bounded operators on $H$ (though typically not producing all of them), this suggests possible generalizations to arbitrary C*-algebras.  This paper is a detailed study of this problem, based on earlier results by R.V. Kadison (1965), F.W. Shultz (1982), K. Thomsen (1982), and others. Perhaps surprisingly, the sharpest known version of Wigner's Theorem for  C*-algebras (which is a variation on a result from Shultz, with considerably simplified proof)
 generalizes the equivalence between the \emph{hypotheses} in the original theorem and those in an analogous result on  (anti-) unitary implementability 
of Jordan automorphisms of $B(H)$, and does not yield (anti-) unitary implementability itself, far from it:  abstract existence results that do give such implementability seem arbitrary and practically useless.
 As such, it would be fair to say that there is no Wigner Theorem for \ca s.
\end{abstract}
\bigskip
\tableofcontents

\thispagestyle{empty}
\renewcommand{\thefootnote}{\arabic{footnote}}
\newpage \setcounter{footnote}{0}
\section{Wigner's Theorem}
In modern notation, Wigner's Theorem reads as follows. Let $H$ be a \Hs\ and let 
\begin{equation}
 \CP_1(H)= \{e\in B(H)\mid e^2=e^*=e, \Tr(e)=\dim(eH)=1\}
\end{equation}
be the set of  one-dimensional projections on $H$, interpreted as the  (normal) pure state space of the C*-algebra $B(H)$; here $\Tr$ is the usual trace. We define a \hi{Wigner symmetry} of $H$ as a bijection  
\beq
\mathsf{W}:\CP_1(H)\raw \CP_1(H) \label{defW}
\eeq
 that satisfies
\begin{equation}
\Tr(\mathsf{W}(e)\mathsf{W}(f))=\Tr(ef), \:\: e,f\in \CP_1(H).\label{UtrUtr}
\end{equation}
\begin{theorem}\label{Wigner}
Each Wigner symmetry of $H$ takes the form 
\begin{equation}
\mathsf{W}(e)= ueu^*, \label{ueu}
\end{equation}
where  $u$ is either unitary or anti-unitary, and is uniquely determined by $\mathsf{W}$ up to a phase.\footnote{This means that $u$ and $u'$  induce the same $\mathsf{W}$ iff $u'=zu$, where $z\in\T$ (i.e.\ the unit circle in the complex plane).} 
\end{theorem}
Condition \er{UtrUtr} states that $\mathsf{W}$ preserves the Born probabilities: if
 $\phv$ and $\psi$ are unit vectors in $H$ such that, in handy Dirac notation,  $e=|\phv\ra\la\phv|$ and $f=|\psi\ra\la\psi|$, then 
$\Tr(ef)=|\la\phv,\psi\ra|^2$, cf.\ \er{epsi}. 

Wigner's Theorem was  first stated by von Neumann and Wigner (1928); see   Bonolis (2004) and Scholz (2006) for some  history. 
 The first  (incorrect) proof  appeared in Wigner (1931), and the first correct one is due to Uhlhorn (1962).  A streamlined version of Wigner's proof,  as improved by  Bargmann (1964), Hunziker (1972), and Simon (1976) may be found in Landsman (2017). See also Cassinelli {\em et al} (1997, 2004),  Chevalier (2007), 
 Keller \emph{et al} (2008), Simon \emph{et al} (2008), and  Freed  (2012) for different approaches. All proofs along these lines are intricate, and, more importantly, they do not seem to hold much promise for generalization to arbitrary \ca s.\footnote{See, though, Moln\'{a}r (1999)  and 
  Baki\'{c} \& Gulja\v{s} (2003)
 for Wigner-type theorems for Hilbert C*-\emph{modules}.}
 
  However, there is a completely different organization of the proof that \emph{does} look promising:
  \begin{enumerate}
\item Prove equivalence between Wigner symmetries of $H$ and what, in honour of Kadison (1965), we call  \hi{Kadison symmetries} of $H$, i.e.\ affine bijections (= isomorphisms of convex sets)
 \beq
 \mathsf{K}: \mathcal{D}(H)\raw \mathcal{D}(H), \label{defK}
 \eeq
where $\mathcal{D}(H)=\{\rho\in B(H)\mid \rho\geq 0, \Tr(\rh)=1\}$ 
 is the convex set of density operators on $H$.
 \item Prove equivalence between Kadison symmetries and what, in honour of Jordan (1933),\footnote{\label{JRC}
The key paper is actually
Jordan, von Neumann, \& Wigner (1934). See also Hanche-Olsen \& St\o rmer (1984). 
 } we call \hi{Jordan symmetries} of $H$, defined as  invertible $\R$-linear bijections
\beq
\mathsf{J}: B(H)_{\mathrm{sa}}\raw B(H)_{\mathrm{sa}}, \label{realJor}
\eeq
 that preserve squares, i.e., $\mathsf{J}(a^2)=\mathsf{J}(a)^2$, or, equivalently, satisfy
 \begin{equation}
\mathsf{J}(a\circ b)=\mathsf{J}(a)\circ\mathsf{J}(b).\label{JPP}
\end{equation}
Here  $B(H)_{\mathrm{sa}}$ is the real vector space of self-adjoint elements of $B(H)$, equipped with the bilinear and commutative but (generally) non-associative \hi{Jordan product} 
\begin{equation}
a\circ b=\half(ab+ba). \label{JP}
\end{equation}
\item Prove (anti-) unitary implementability of Jordan symmetries of $H$.
\end{enumerate}
\newpage
\noindent Note that none of these three symmetry concepts requires any topology and continuity.
This program works: see  Bratteli \& Robinson (1987), \S 3.2.1, Moln\'{a}r (1998), Moretti (2013), Chapter 12, or Landsman (2017), Chapter 5.
The three steps above are carried out as follows:
\begin{proposition}\label{Eqall}
\begin{enumerate}
\item There is an isomorphism  $\mathsf{W}\lraw \mathsf{K}$ of groups between the group of Wigner symmetries \er{defW} of $H$
and the group of Kadison symmetries \er{defK} of $H$ such that
\begin{equation}
\mathsf{W}=\mathsf{K}_{| \CP_1(H)}.\label{UsfU1}
\end{equation}
\item There is an anti-isomorphism  of groups  $\mathsf{K}\lraw \mathsf{J}$ between the group of Kadison symmetries \er{defK} of $H$
and the group of Jordan symmetries \er{realJor} of $H$ such that
\begin{equation}
\Tr(\mathsf{K}(\rho)a)=\Tr(\rh\mathsf{J}(a)) \:\:\: \mathrm{for\:\: all }\:\:\: \rh\in\mathcal{D}(H),\, a\in B(H)_{\mathrm{sa}}. \label{JK}
\end{equation}
\item Each Jordan symmetry of $H$ takes either one of the following two forms:
\begin{align}
\mathsf{J}_{\C}(a)&= u^*au \hspace{50pt} (u\: \mbox{unitary}); \label{uau1} \\
\mathsf{J}_{\C}(a)&= u^*a^*u  \hspace{45pt}  (u\: \mbox{anti-unitary}), \label{uau2} 
\end{align}
where  $u$ is uniquely determined by $\mathsf{J}$ up to a phase. The first case holds iff $\J_{\C}:B(H)\raw B(H)$ is an \emph{automorphism} of \ca s, whereas in the second case it is an
\emph{anti-automorphism}.
\end{enumerate}
\end{proposition}
Here we have stated point 3 in terms of the complexification $\J_{\C}$ of $\J$ as defined in 
\er{realJor}; in terms of $\J$ eq.\ \er{uau2} would read $\J(a)=u^*au$, which hides what is really going on.
Any $\R$-linear map $\phv:B(H)_{\mathrm{sa}}\raw B(H)_{\mathrm{sa}}$
 has a unique $\C$-linear *-preserving extension $\phv_{\C}: B(H) \raw B(H)$, defined by
 $\phv_{\C}(a+ib)=\phv(a)+i\phv(b)$, where $a^*=a$ and $b^*=b$.
Conversely,  a  $\C$-linear *-preserving map $\phv_{\C}$
defines a  map $\phv$ by restriction to $B(H)_{\mathrm{sa}}$. Thus we may work with either $\J$ or $\J_{\C}$, and with abuse of notation we often even write  $\J$ for $\J_{\C}$, where the context indicates the real or complex setting.

Detailed proofs of the first two parts may be found in Landsman (2017),  Propositions 5.15 and 5.19, respectively.
The third part is Theorem \ref{step3} below,  based on the highly  intricate Lemma \ref{Thomsen}.

 Proposition \ref{Eqall} implies Theorem \ref{Wigner}, since parts 1 and 2 imply:
 \begin{corollary}\label{EqWigJor}
There is an anti-isomorphism of groups $\mathsf{W}\lraw \mathsf{J}$ between the group of Wigner symmetries \er{defW} of $H$
and the group of Jordan symmetries \er{realJor} of $H$, such that
\begin{equation}
\Tr(\mathsf{W}(e)a)=\Tr(e\mathsf{J}(a)) \:\:\: \mathrm{for\:\: all }\:\:\: e\in\CP_1(H),\, a\in B(H). \label{JW}
\end{equation}
\end{corollary}
Part 3 and the cyclic property of the trace then give, initially for $a^*=a$ and hence in general,\footnote{For $a\in B(H)$,
in the  anti-unitary case \er{uau2} one uses the property $\la u\psi, u\phv\ra=\ovl{\la\phv,\psi\ra}$ 
and hence $\Tr(a)=\ovl{\Tr(uau^*)}$, with $uu^*=1_H$,
to compute $\Tr(\mathsf{W}(e)a)=\Tr(eu^*a^*u)=\overline{\Tr(ueu^*a^*uu^*)}=\overline{\Tr(ueu^*a^*)}=
\Tr(aueu^*)=\Tr(ueu^*a)$.}
\begin{equation}
\Tr(\mathsf{W}(e)a)=\Tr(e\mathsf{J} (a))=\Tr(eu^*au)=\Tr(ueu^*a),
\end{equation}
upon which non-degeneracy of the trace pairing yields \er{ueu}. For the record, we also state:
\begin{corollary}
Each Kadison symmetry of $H$ takes the form 
\begin{equation}
\mathsf{K}(\rh)= u\rh u^*, \label{uau3}
\end{equation}
where  $u$ is either unitary or anti-unitary, and is uniquely determined by $\mathsf{K}$ up to a phase.
\end{corollary} 
Running the argument backwards,  this corollary is equivalent to Wigner's Theorem \ref{Wigner} as well as to 
Proposition \ref{Eqall}.3. Furthermore, there are at least three other Wigner-like theorems equivalent to any of the three versions just given (Landsman, 2017, Chapter 5): each states that any automorphism of some specific structure related to a \Hs\ $H$ is  implemented (anti-) unitarily.\footnote{These structures are von Neumann's \emph{projection lattice} $ \CP(H)= \{e\in B(H)\mid e^2=e^*=e\}$, the \emph{effect algebra}  $\CE(H)=\{a\in B(H)\mid 0\leq a\leq 1_H\}$, and the \emph{Bohr poset} $\CC(B(H))=\{C\subset B(H)\mid C\mbox{ commutative \ca}, 1_H\in C\}$.}
 However, for the current paper pure states, mixed states, and (Jordan) operator algebras will do.
\section{Generalizing \tpies}\label{GTP}
The question, then, is if and how Wigner's Theorem \ref{Wigner} may be generalized to arbitrary \ca s.\footnote{We assume familiarity with the basic theory of \ca s at the level of books like Bratteli \& Robinson (1987), Dixmier (1977), Kadison \& Ringrose (1983), or Landsman (2017), Appendix C. 
} It is  clear what pure states are in general, but Theorem \ref{Wigner} does not use pure states on $B(H)$ but \emph{normal} pure states. Similarly, Kadison symmetries of $H$ act on the \emph{normal} state space $S_n(B(H))\cong\mathcal{D}(H)$ rather than the state space $S(B(H))$. This problem will be resolved through appropriate continuity conditions on symmetries, 
see \S\ref{GWT} below.\footnote{A Wigner Theorem for normal pure states on von Neumann algebras would be self-defeating, since many von Neumann algebras (ranging from  $L^{\infty}(0,1)$ to type {\sc ii} and type {\sc iii} factors) have no normal \emph{pure} states. In a fascinating handwritten manuscript from 1937 that was only published 44 years later, von Neumann (1981) investigates \tpies\  $P:\CP(M)\x \CP(M)\raw[0,1]$ on the lattice $\CP(M)$ of \emph{all} projections in a  type {\sc ii}$\mbox{}_1$ factor $M$, where  $P(e,f)$ is interpreted as the \tp\ \emph{from} $e$ \emph{to} $f$. If $\mathrm{tr}:M\raw\C$ is the unique finite trace  normalized as $\mathrm{tr}(1_M)=1$, with associated dimension function $d:\CP(M)\raw[0,1]$ simply defined as $d(e)=\mathrm{tr}(e)$, 
  one has $P(e,f)=d(f)/ d(e)$ whenever $f\leq e$ and $e\neq 0$.  Von Neumann interpreted $d(e)$ as the \emph{a priori probability} of $e$, on which the \emph{\tp} $P(e,f)$ of $f$ relative to $e$ is predicated. This interpretation obviously does not work for the type {\sc i} factor $B(H)$ for infinite-dimensional $H$, since $\mathrm{Tr}(e)=\infty$ for any infinite-dimensional projection $e$, which is why, according to R\'{e}dei (1996), von Neumann felt that in \qm\  type {\sc ii}$\mbox{}_1$ factors should replace $B(H)$.
The theory of von Neumann (1981) is very interesting, but since his \tp\ $P$ is neither symmetric nor satisfies the condition $P(e,f)=1$ iff $e=f$, it has a rather different flavour from the one described in the present paper. One could also work with the symmetric bilinear expression 
$\sg(e,f)=\mathrm{tr}(ef)$ on $\CP(M)\x \CP(M)$, about which Moln\'{a}r (2000) proved that any bijection of $\CP(M)$ that preserves $\sg$ is given by 
 either an automorphism or an anti-automorphism  $\mathsf{J}$ of $M$ (and hence by a Jordan automorphism). However, in general 
  $\mathsf{J}$ cannot be (anti-) unitarily implemented, cf.\ Connes (1975). 
\label{vNfn1}
} Granting this, we now ask what a \tp\ on the pure state space $P(A)$ of an arbitrary \ca\ $A$ should be.

 This question touches on the essence of the historical development of quantum theory, since besides providing a rigorous proof of the mathematical equivalence between Heisenberg's matrix mechanics and Schr\"{o}dinger's wave mechanics (in the footsteps of Pauli, Dirac, and Schr\"{o}dinger himself), the other main goal of  von Neumann's (1932) 
 axiomatization of \qm\ was precisely the search for a mathematical home for the mysterious transition amplitudes 
$\la\psi|\phv\ra$, 
 and ensuing \tpies\ $|\la\psi|\phv\ra|^2$   introduced in 1926--1927 by Born, Jordan, and Dirac.\footnote{
  The literature on von Neumann (1932) is huge. See e.g.\ Birkhoff \&  Kreyszig (1984), R\'{e}dei \& St\"{o}ltzner (2001), 
  Duncan \& Janssen (2013), Dieks (2016), and Landsman (2019).} 

 Von Neumann 
 found this home in \Hs: if the symbols $|\psi\ra$ and $|\phv\ra$ are identified with unit vectors $\psi$ and $\phv$ in some \Hs\ $H$, then $\la\psi|\phv\ra$ is simply the inner product $\la\psi,\phv\ra$ of $\psi$ and $\phv$ (which physicists even write as $\la\psi|\phv\ra$, but this blurs the fundamental conceptual difference between the physical transition amplitudes $\la\psi|\phv\ra$ and their mathematical identification with inner products). Von Neumann was one of the first to recognize that pure states  are not unit vectors $\psi$ but equivalence classes therefore (namely up to phase factors), whose equivalence class is best seen as the corresponding one-dimensional projections $e_{\psi}=|\psi\ra\la\psi|$, so that, where the transition \emph{amplitudes} $\la\psi|\phv\ra$ unsatisfactorily depend  on the phases of $\psi$ and $\phv$, the corresponding transition \emph{probabilities} do not, for they can be written directly in terms of the projections by
 \begin{equation}
 |\la\psi|\phv\ra|^2=\Tr(e_{\psi}e_{\phv}), \label{epsi}
\end{equation}
and so the \tp\ $P(e,f)$ between pure states $e,f\in\CP_1(H)$ is \emph{intrinsically} given by
\begin{equation}
P(e,f)=\Tr(ef). \label{tref}
\end{equation}
However, even at the time this did not fully capture what had been proposed by the physicists: apart from quantum systems with infinitely many degrees of freedom (for which von Neumann would later develop a possible mathematical basis in the form of operator algebras), his formula for the \tp\ only made sense for vectors in \Hs\ (or projections thereon), whereas Dirac and Jordan also heavily used 
probability amplitudes and the ensuing
\tpies\ between ``eigenstates" $|\lm\ra$ for continuous eigenvalues $\lm$ (i.e.\ lying in the continuous spectrum of some physical observable). 
A typical example is the ubiquitous expression 
\begin{equation}
\la x|p\ra=e^{-ixp/\hbar}, \label{exp}
\end{equation}
where $x,p\in\R$, which has no counterpart in von Neumann's (1932) \Hs\ formalism.\footnote{Von Neumann (1932)  tried to capture \er{exp} by expressions like $\Tr(E(I)F(J))$, where $E(I)$ and $F(J)$ are the spectral projections for subsets $I$ and $J$ in the spectra of some self-adjoint operators $A$ and $B$, respectively. Unfortunately, these ``\tpies" are not only 
unnormalized; they may even be infinite. Von Neumann (1961)
 sought a way out of this problem through his  discovery of type  {\sc ii}$\mbox{}_1$ factors,  cf.\ footnote \ref{vNfn1}. Indeed, 
replacing $\Tr(E(I)F(J))$ by $\mathrm{tr}(E(I)F(J))$ then makes the \tpies\ finite as well as normalized. See Araki (1990) and R\'{e}dei (1996, 2001). \label{RFN}
} Indeed, the only reasonable interpretation is that the corresponding transition probability should be zero, rather than one, as  (absurdly) suggested by taking the square of the absolute value of \er{exp}.

We may reinterpret both \er{tref} and the above consideration from the point of view of \ca s $A$,  replacing the non-available language of non-normal states (which would be one way to look at Dirac states like $|x\ra$ and $|p\ra$) by the available language of inequivalent GNS-\rep s. We say that two states  $\om$ and $\om'$ on $A$ are \emph{equivalent} iff the corresponding {\sc gns}-\rep s $\pi_{\om}$ and $\pi_{\om'}$ are unitarily equivalent, and  \emph{inequivalent} otherwise. In the first case, by Theorem 5.19 (vi) in Alfsen \& Shultz (2001) states that there is a single irreducible \rep\ $\pi:A\raw B(H)$ such that $H$ contains two unit vectors $\Om_{\om}$ and $\Om_{\om'}$, each cyclic for $\pi(A)$, such that for all $a\in A$,
\begin{align}
\om(a) = \la \Om_{\om}, \pi(a)\Om_{\om}\ra; && \om'(a) = \la \Om_{\om'}, \pi(a)\Om_{\om'}\ra. \label{oma}
\end{align}
Moreover, the number $|\la \Om_{\om},\Om_{\om'}\ra|$ only depends on $\om$ and $\om'$ (and not on the choice of $\pi$ etc.).
\begin{definition}\label{tpCAnorm}
Let $A$ be \ca\ with pure state space $P(A)$. The \emph{\tp} 
\begin{equation}
\ta:P(A)\x P(A)\raw [0,1]
\end{equation}
is defined by the following two cases, where $\om,\om'\in P(A)$:
\begin{enumerate}
\item If $\om$ and $\om'$ are equivalent, in which case we may assume \er{oma},
 we define
\begin{equation}
\ta(\om,\om')= |\la \Om_{\om},\Om_{\om'}\ra|^2. \label{tpeqstates}
\end{equation}
\item  If $\om$ and $\om'$ are inequivalent, we define
 \begin{equation}
\ta(\om,\om')=0. \label{tpineqstates}
\end{equation}
\end{enumerate}
\end{definition}
Even the case distinction can be avoided, since by Landsman (2017), Proposition C.177, we have
\begin{equation}
\ta(\om,\om')=1-\quar\|\om-\om'\|^2, \label{tp22}
\end{equation}
as in Roberts \& Roepstorff (1969).  Also, from Landsman (1998), Theorem I.2.8.2 we may write
\begin{equation}
\ta(\om,\om')= \inf\{\om(a)\mid a\in A, 0\leq a\leq 1_A,\om'(a)=1\},\label{tpCstarA}
\end{equation}
see also  Landsman (2017), eq.\ (C.629). Yet another expression for $\ta$, used by Shultz (1982), is
\begin{equation}
\ta(\om,\om')=\tilde{\om}(\mathrm{carrier}(\om'))=\tilde{\om'}(\mathrm{carrier}(\om)), \label{tpsh}
\end{equation}
where  $\tilde{\om}\in S_n(A^{**})$ is the unique extension of $\om\in S(A)$ to a normal state on its double dual $A^{**}$, seen as a von Neumann algebra as usual, and $\mathrm{carrier}(\om)=\inf\{e\in\CP(A^{**})\mid \til{\om}(e)=1\}$ is the carrier  (or support projection) of $\om$ in $A^{**}$. The equivalence between \er{tpsh} and \er{tp22} will be proved in 
Lemma \ref{RT3.79} below. The symmetry of $\ta$ in its arguments is clear from \er{tpeqstates}, \er{tpineqstates}, and  \er{tp22}, 
and hence follows also for \er{tpsh}. This makes $\ta$ a transition probability in an axiomatic sense going back to Mielnik (1968), see also Landsman (1998), but this will not be used here.
\section{Generalizing Wigner's Theorem}\label{GWT}
As first recognized by Kadison (1965) for what we call Kadison symmetries of the total state space,  for general \ca s one needs some (uniform) continuity assumptions.\footnote{Much as topological spaces were invented to express the idea of continuity, \emph{uniform} spaces were invented in order to generalize the idea of \emph{uniform} continuity (from metric spaces to more general settings). See Bourbaki (1989), Chapter II, for what follows and more. 
 A \hi{uniform structure} or \hi{uniformity} on a set $X$ is a nonempty filter  $\mathcal{U}$ on $X\x X$ 
(i.e.,a collection $\mathcal{U}\subset\CP(X\x X)$ of subsets $U$ of $X\x X$, called \hi{entourages}, playing the role of open sets in topology, such that  $U\in \mathcal{U}$ and $U\subset V$ imply $V\in \mathcal{U}$, 
and $U\in \mathcal{U}$ and $V\in \mathcal{U}$ imply $U\cap V\in \mathcal{U}$) satisfying the following conditions:\\
1.  Each $U\in \mathcal{U}$ contains the diagonal $\Dl_X=\{(x,x)\mid x\in X\}$;
2. If $U\in \mathcal{U}$, then  $U^T\in \mathcal{U}$, where $U^T=\{(y,x)\mid (x,y)\in U\}$;
3. If $U\in \mathcal{U}$, then there is some $V\in \mathcal{U}$ such that $V^2\subseteq U$, where
$V^2=\{(x,z)\mid \exists y\in X:(x,y)\in V, (y,z)\in V\}$.

A set with a uniformity is called a \hi{uniform space}. Each uniformity on $X$ generates a topology on $X$ for which 
the sets $V_x=\{y\in X\mid (x,y)\in \mathcal{U}\}$ form a neighbourhood filter at $x$.
There is a natural notion of a fundamental system of entourages generating a uniformity. For example, a metric space $(X,d)$ carries a uniformity generated by sets $U_{\varep}=\{(x,y)\in X\x X\mid d(x,y)<\varep\}$, where $\varep>0$, whose associated topology is the metric topology. 
If $X$ and $Y$ are uniform spaces, a function $f:X\raw Y$ is \hi{uniformly continuous} if $f\inv (V)\in \mathcal{U}_X$ whenever $V\in \mathcal{U}_Y$. A uniformly continuous is automatically continuous for the topologies defined by the uniformities. For the main text the following fact is important, see Theorems I and II in \S II.4.1. of  Bourbaki (1989):
A \emph{compact} space $X$ carries a unique uniformity whose associated topology is the given one (whose entourages are all neighbourhoods of the diagonal in $X\x X$), and any continuous function from $X$ to a uniform space is uniformly continuous (and \emph{vice versa}, as already noted). \label{Bourbaki}} In this context the total state space $S(A)$ of a \ca\ $A$, for simplicity assumed to be unital,\footnote{In all results the non-unital case is reduced to the unital case by unitization, as explained in detail in Rang (2019).} naturally carries the $w^*$-\emph{topology} as well as the closely related (and crucial) $w^*$-\emph{uniformity}.\footnote{Let $A$ be our unital \ca. Since the state space $S(A)$ is compact in the $w^*$-topology, it carries a unique uniformity compatible with this topology in the sense just explained. This uniformity is simply the restriction of the canonical $w^*$-unformity $\mathcal{U}_{w^*}$ on $A^*$, which is generated by subsets  $\{(\phv,\phv')\in A^*\x A^*: |\phv(a)-\phv'(a)|<\varep\}$, where $a\in A$ and $\varep>0$. By the theorems in footnote \ref{Bourbaki}, any function $f:S(A)\raw S(A)$ is continuous iff it is uniformly continuous. By Proposition \ref{KadA1}, if an affine bijection  \er{defKA1} is a pullback of a Jordan symmetry, then it must be $w^*$-continuous, and hence  uniformly continuous. Therefore, if  Proposition \ref{KadA2}, Theorem \ref{Kitty}, and Theorem \ref{Shultz} have any hope of being true, 
in combination with Proposition \ref{KadA1} it follows that  Wigner symmetries \er{defWA} and $\CS$-Kadison symmetries \er{defKA2} are necessarily restrictions of Kadison symmetries \er{defKA1}, and hence they are not only $w^*$-continuous but also  uniformly continuous. Since  $\CS(A)$ and $P(A)$ are non-compact subspaces of $S(A)$, for maps  \er{defWA} and \er{defKA2} continuity and uniform continuity are not \emph{a priori} equivalent, 
and the stronger assumption of uniform continuity is necessary (and sufficient). \label{wstar}
}
 For the pure state space  $P(A)$ one requires the  $w^*$-{uniformity} in any case (Shultz, 1982). We will also encounter the real part  $A_{\mathrm{sa}}=\{a\in A\mid a^*=a\}$ of $A$,  seen as a Jordan algebra under the product \er{JP}.
\begin{definition}\label{defAWKJ}
Let $A$ be a \ca\ with total state space $S(A)$, equipped with the $w^*$-topology, and 
pure state space $P(A)$, equipped with the relative $w^*$-uniformity (both inherited from $A^*$).
\begin{enumerate}
\item 
 A \hi{Wigner symmetry} of $A$ is a uniformly continuous bijection 
\begin{equation}
\mathsf{W}: P(A)\raw P(A) \label{defWA}
\end{equation}
with uniformly continuous inverse
that preserves \tpies\ in the sense that
\begin{equation}
\ta(\mathsf{W}(\om), \mathsf{W}(\om'))=\ta(\om,\om'),
\end{equation}
for all $\om,\om'\in P(A)$,
where the function $\ta:P(A)\x P(A)\raw [0,1]$ is given by Definition \ref{tpCAnorm}.
\item A  \hi{Kadison symmetry} of $A$ is an affine homeomorphism
\begin{equation}
\mathsf{K}: S(A)\raw S(A). \label{defKA1}
\end{equation}
\item A \hi{Jordan symmetry} of $A$ is an invertible $\R$-linear bijection
\beq
\mathsf{J}: A_{\mathrm{sa}}\raw A_{\mathrm{sa}}. \label{realJorA}
\eeq
 that  satisfies \er{JPP}, or, equivalently, preserves squares,  i.e.\ $\mathsf{J}(a^2)=\mathsf{J}(a)^2$.
\end{enumerate}
\end{definition}
Proposition \ref{Eqall}.2 then has an obvious 
 generalization (Alfsen \& Shultz, 2001, Corollary 4.20):
\begin{proposition}\label{KadA1}
For any \ca\ $A$
there is an anti-isomorphism of groups $\mathsf{K}\lraw \mathsf{J}$ between the group of Kadison symmetries \er{defKA1} of $A$
and the group of Jordan symmetries \er{realJorA} of $A$, such that
\begin{equation}
\mathsf{K}(\om)=\om\circ \mathsf{J}, \:\:\: \om\in S(A). \label{KJA}
\end{equation}
\end{proposition}
This should be regarded as elementary and is proved in  exactly the same way as Proposition \ref{Eqall}.2, namely by first establishing a bijective correspondence between Jordan symmetries of $A$ and positive unital maps $A\raw A$, and subsequently between those  and Kadison symmetries of $A$.

 Perhaps partly because  Proposition \ref{KadA1} does not reduce to Proposition \ref{Eqall}.2 if $A= B(H)$, Kadison (1965)  proved a refinement of Proposition \ref{KadA1}, based on the concept of a \hi{full} family of states. This is a convex subset
$\mathcal{S}(A)$ of the total state space $S(A)$ that is $w^*$-dense, or, equivalently,\footnote{This equivalence is nontrivial; see Kadison (1965), Theorem 2.2, or  Bratteli \& Robinson (1987), Proposition 3.2.10. Note that what Kadison (curiously) calls a C*-automorphism is a Jordan automorphism in our (and the usual) language.
}
 is such that $a\geq 0$ iff $\om(a)\geq 0$ for all $\om\in \mathcal{S}(A)$. 
 For example, the density matrices $\mathcal{D}(H)$ define a full family of states on $A=B(H)$ via the affine bijection  $\mathcal{D}(H)\cong S_n(B(H))$ given by $\rho\lraw\om$ with $\om(a)=\Tr(\rh a)$, where $S_n(B(H))$ is the normal state space of $B(H)$, seen as a von Neumann algebra. The fact that  $S_n(B(H))$ is full in $S(B(H))$ easily follows
 from the second criterion for fullness, which indeed is already satisfied by vector states $\om(a)=\la\psi,a\psi\ra$, where $\psi\in H$ is a unit vector; note that $\CP_1(H)\subset \mathcal{D}(H)$. This leads to the following refinement of Definition \ref{defAWKJ}.2:
\begin{definition}\label{CSA}
Let $A$ be a \ca\ with full family of states $\mathcal{S}(A)$, equipped with the relative $w^*$-uniformity from $A^*$.
A  \hi{$\CS$-Kadison symmetry} of $A$ is a uniformly continuous affine bijection 
\begin{equation}
\mathsf{K}: \CS(A)\raw \CS(A) \label{defKA2}
\end{equation}
with uniformly continuous inverse.
\end{definition}
  $\CS$-Kadison symmetries suffice to recover Jordan symmetries, for according to Kadison (1965), Theorem 3.3 (also cf.\ Theorem 3.2.11 in Bratteli \& Robinson, 1987), Proposition \ref{KadA1} refines to:
\begin{proposition}\label{KadA2}
For any \ca\ $A$
there is an anti-isomorphism of groups $\mathsf{K}\lraw \mathsf{J}$ between the group of $\CS$-Kadison symmetries \er{defKA2} of $A$
and the group of Jordan symmetries \er{realJorA} of $A$, s.\ t.
\begin{equation}
\mathsf{K}(\om)=\om\circ \mathsf{J}, \:\:\: \om\in \CS(A). \label{KJA2}
\end{equation}
\end{proposition} 
If $A=B(H)$ and $\CS(A)=S_n(B(H))$ this  recovers Proposition \ref{Eqall}.2 (which Proposition \ref{KadA1} fails to do). Otherwise, we will not need Definition \ref{CSA} and Proposition \ref{KadA2}, except for Shultz's  idea of using the $w^*$ uniformity on $P(A)$. Our main result is the following  generalization of Corollary \ref{EqWigJor}:
\begin{theorem}\label{Kitty}
For any \ca\ $A$
there is an anti-isomorphism of groups $\mathsf{W}\lraw \mathsf{J}$ between the group of Wigner symmetries \er{defWA} of $A$
and the group of Jordan symmetries \er{realJorA} of $A$, such that
\begin{equation}
\mathsf{W}(\om)=\om\circ \mathsf{J}, \:\:\: \om\in P(A). \label{JWA}
\end{equation}
\end{theorem} 
As far as we know, this theorem has not appeared in the literature, although it is closely related to Theorem \ref{Shultz} below from Shultz (1982) . Theorem \ref{Kitty} and Propositions \ref{KadA1} and \ref{KadA2} imply
\begin{corollary}
For any \ca\ $A$ there is an isomorphism  $\mathsf{W}\lraw \mathsf{K}$ of groups between the group of Wigner symmetries \er{defWA} of $A$
and  the group of either Kadison symmetries \er{defKA1} or  $\CS$-Kadison symmetries \er{defKA2} 
of $A$, such that for either all $\om\in P(A)$ or all $\om\in\CS(A)\cap\,  P(A)$ one has
\begin{equation}
\mathsf{W}(\om)=\mathsf{K}(\om). \label{UsfU1A}
\end{equation}
\end{corollary}
This is obviously  a generalization of Proposition \ref{Eqall}.1, but the order of things is different between $B(H)$ and the case of general \ca s $A$: for $B(H)$ one first proves the equivalence between Wigner  and Kadison symmetries, and independently between Kadison and Jordan symmetries, to conclude the equivalence between Wigner and Jordan symmetries (which in turn is the main lemma in this specific approach to Wigner's Theorem), whereas for general $A$ one independently proves the equivalences between 
 Wigner and Jordan symmetries and between Jordan and Kadison symmetries, to conclude the equivalence between Wigner and Kadison symmetries.
\begin{corollary}
 Wigner and Kadison symmetries of a \Hs\ $H$ are uniformly continuous.
\end{corollary}
So far, we have no generalization of Wigner's Theorem \ref{Wigner}. Both in the interest of finding one (albeit in vain) and as the key lemma for step 3 in the process explained in the Introduction, i.e.\ the (anti-) unitary implementability of Jordan symmetries of $H$, we adapt, from Thomsen (1982):
\begin{lemma}\label{Thomsen}
Let $\mathsf{J}: A_{\mathrm{sa}}\raw A_{\mathrm{sa}}$ be a Jordan symmetry of a 
concretely represented \ca\ $A\subset B(H)$, with complexification $\mathsf{J}_{\C}: A\raw A$.
 There  exist   projections $p_1$, $p_2$, and  $p_3$ in the center of $A''$ that satisfy the following conditions and are are uniquely determined by these conditions:
 \begin{enumerate}
 \item  $p_1$, $p_2$, and $p_3$ are mutually orthogonal (i.e.\ $p_1p_2=p_1p_3=p_2p_3=0$);
\item $p_1+p_2+p_3=1_H$;
\item The map $a\mapsto \mathsf{J}_{\C}(a)p_1$  is a homomorphism and not
an anti-homomorphism (of \ca s);
\item The map $a\mapsto \mathsf{J}_{\C}(a)p_2$ is an anti-homomorphism and not a homomorphism;
\item The map $a\mapsto \mathsf{J}_{\C}(a)p_3$ is both  a homomorphism and an anti-homomorphism.\footnote{Here a 
homomorphism is a complex-linear map $\phv:A\raw B(H)$ that satisfies $\phv(a^*)=\phv(a)^*$ and $\phv(ab)=\phv(a)\phv(b)$.
 An anti-homomorphism is a  complex-linear map $\phv:A\raw B(H)$ that satisfies  $\phv(a^*)=\phv(a)^*$ and $\phv(ab)=\phv(b)\phv(a)$. What many authors call a  $\mbox{}^*$-homomorphism (or $\mbox{}^*$-isomorphism) is simply  a homomorphism (etc.) in our terminology.}
\end{enumerate}
\end{lemma} 
See \S\ref{herstein} for a detailed proof, or
Proposition 5.69 in Alfsen \& Shultz (2001), whose approach is quite different. 
 In this situation one really obtains three different complex-linear maps 
 \begin{align}
\mathsf{J}_i: A&\raw B(p_iH), \:\:\: i=1,2,3,\\
\mathsf{J}_i(a)&=p_i \mathsf{J}_{\C}(a)p_i,
\end{align}
of which $\mathsf{J}_1$ and $\mathsf{J}_3$ as well as their direct sum are homomorphisms, and $\mathsf{J}_2$ and $\mathsf{J}_3$ as well as their direct sum are anti-homomorphisms. Note that $\mathsf{J}_3(A)$ must be commutative in $B(p_3H)$. 
\begin{theorem}\label{step3}
A Jordan symmetry $\J$ of $B(H)$ is either an automorphism, in which case $\J(a)=u^*au$ for some unitary $u$, 
or an anti-automorphism, in which case $\J(a)=u^*a^*u$ for some anti-unitary $u$.
\end{theorem}
\emph{Proof.} The center of $A=B(H)$ is trivial. If $H\cong\C$ we have $p_1=p_2=0$ and $p_3=1$, which case is almost trivial, so assume $\dim(H)\geq 2$. Then $p_3=0$ and  either $p_1=1_H$ and hence $p_2=0$, in which case  $\mathsf{J}$ is an automorphism, or $p_1=0$ and hence $p_2=1_H$, in which case $\mathsf{J}$ is an anti-automorphism. 

We only treat the former case, to which the latter is easily reduced, cf.\ Proposition 5.28 in Landsman (2017).\footnote{The following  argument comes from Bratteli \& Robinson (1987), Example 3.2.14, or Landsman (2017), \S 5.4.}
 If  $\al:B(H)\raw B(H)$ is an automorphism and  $e\in B(H)$ is a one-dimensional projection,  then so is $\al(e)$, see Lemma 5.27 in  Landsman (2017). Fix a unit vector $\chi\in H$, with  one-dimensional projection $e_{\chi}=|\chi\ra\la\chi|$, 
and choose  a  unit vector $\phv$ such that $e_{\phv}=\al\inv(e_{\chi})$; given $\chi$, this determines $\phv$ up to a phase. Any $\psi\in H$ can be written as $\psi=b\chi$ for some $b\in B(H)$, and hence also as $\psi=\al(a)\chi$, for some $a\in B(H)$. We attempt to define an operator $u$ by $u\al(a)\chi=a\phv$.
  
\noindent Although different operators $a$  give rise to the same vector $\al(a)\chi$, in fact $u$ is well defined, since
\begin{align}
\|u\al(a)\chi\|_H&=\|a\phv\|_{H}=\| a e_{\phv}\|_{B(H)}=\|\al(a e_{\phv})\|_{B(H)}\nn
\\ &=\|\al(a)\al(e_{\phv})\|_{B(H)}=\|\al(a)e_{\chi}\|_{B(H)}=\|\al(a)\chi\|_H, \label{3.12}
\end{align}
since $\al$ is an isometry. Hence if $\al(a)\chi=\al(b)\chi$, then $u\al(a)\chi=u\al(b)\chi$.
Clearly $u$ is isometric by \er{3.12}, and since it is also surjective (because $\{a\phv\mid a\in B(H)\}=H$ for any $\phv\neq 0$), it is unitary. 
  
 Next, by definition of $u$ we have $u\al(ab)\chi=ab\phv$, and 
  since $\al$ is an automorphism this implies $u\al(a)\al(b)\chi=ab\phv$. With $\al(b)\chi=u^*b\phv$, this reads
  $u\al(a)u^*b\phv=ab\phv$, and since $b\in B(H)$ is arbitrary and hence $b\phv\in H$ is arbitrary, this gives
    $u\al(a)u^*=a$ and hence $\al(u)=u^*au$, as claimed. 

Finally, if for all $a\in B(H)$ we have $u^*au=v^*av$ for unitaries $u$ and $v$, then, using $uu^*=1_H$ and $vv^*=1_H$ we obtain $[a,uv^*]=0$, so that by Schur's Lemma
$uv^*=\lm\cdot 1_H$ for some $\lm\in\C$, which unitary forces into $\lm\in\T$. 
As explained in the Introduction, Wigner's Theorem \ref{Wigner} follows. \hfill $\Box$
\smallskip

Also for general \ca s $A$, if there is to be any hope for nontrivial (anti-) unitarily implementability of $\mathsf{W}$  \`{a} la \er{ueu}, then either of the projections $p_1$ or $p_2$ should vanish. To enforce this, one may add a further condition on the Wigner symmetry $\mathsf{W}$, namely that it be either \hi{orientation-preserving} or  \hi{orientation-reversing}. This refers to a further structure on $P(A)$ introduced by Shultz (1982), which we explain in \S\ref{shultz} below. 
Theorem \ref{Kitty} may then be refined as follows:\footnote{In the orient\-ation-preserving case this duly induces an anti-isomorphism of groups, as in Theorem \ref{Kitty},
but the orientation-reversing symmetries do not form a group: the composition of two of these is  orient\-ation-preserving.}
\begin{theorem}\label{Shultz}
For any \ca\ $A$ there is a bijective correspondence $\mathsf{W}\lraw \al$ between orient\-ation-preserving (reversing)  Wigner symmetries $\W$ of $A$ and (anti-) automorphisms $\al$ of $A$, s.t.
\begin{equation}
\mathsf{W}(\om)=\om\circ \mathsf{\al}, \:\:\: \om\in P(A). \label{JWA2}
\end{equation}
\end{theorem} 
 This is (essentially) Theorem 18 in Shultz (1982), which is neither weaker nor stronger than our Theorem \ref{Kitty}; clearly its conclusion is stronger, but so are its assumptions.
If $\mathsf{W}$ is orientation-preserving, then in Lemma \ref{Thomsen} the symmetry $\mathsf{J}=\al$ is a homomorphism of \ca s, and hence $p_2=0$. To find a unitary even in  that favourable case we need Lemma 4.12 in Kadison (1965), which invokes another \Hs\ $H'$. A ``Wigner Theorem" within reach would  be:
\smallskip

\noindent\emph{For any
\ca\ $A$ acting on a \Hs\ $H$  there is a \Hs\ $H'$ such that for
 any  orientation-preserving Wigner symmetry $\mathsf{W}$ of $A$ there is a unitary operator $u$ on $H\otimes H'$ satisfying }
\begin{equation}
\al(a)\ot 1_{H'}=u(a\ot 1_{H'})u^*, \label{K57}
\end{equation}
\emph{where $\al\in \mathrm{Aut}(A)$ is the automorphism of $A$ determined by $\mathsf{W}$ according to Theorem \ref{Shultz}.}
  \smallskip

\noindent
However, this ``theorem" is quite useless: the choice of $H$ on which $A$ is represented concretely is usually arbitrary (except when $A=B(H)$), the construction of $H'$ is awful (Kadison, 1957, Lemma 4.1.7), and consequently nothing suggests that $u$ is uniquely determined by $\mathsf{W}$. Flogging a dead horse, one cannot formulate the statement of the theorem directly in terms of $\mathsf{W}(\om)$, as in \er{ueu}, since that would necessitate yet another arbitrary choice, namely of an extension of $\om\in P(A)$ to some (pure?) state om $B(H_1\ot H_2)$. 
Compare this with the elegance of Wigner's Theorem \ref{Wigner}! 

Our conclusion is that, of the three-step program for proving Theorem \ref{Wigner} for $A=B(H)$, for general \ca s $A$ only a fusion of the first two steps is canonical, leading to Theorem \ref{Kitty}, whereupon any further steps towards mimicking the original version do not make enough sense to be carried out. In this respect, as far as we know \emph{there simply is no Wigner Theorem for general \ca s}. We have to satisfy ourselves with Theorem \ref{Kitty}, which  in the context of the original theorem would merely comprise a step in the proof, and with its refinement Theorem \ref{Shultz}.

Theorem \ref{Kitty} and Theorem \ref{Shultz} are proved in sections \ref{kitty} and \ref{shultz}, respectively.
We prove the second from the first and as such our proof is quite different from the proof in
Shultz (1982).
\section{Proof of main theorem}\label{kitty}
In this section we prove Theorem \ref{Kitty}. The easy part of the proof assigns a Wigner symmetry $W$ as defined in  \er{defWA} to a Jordan symmetry $\J$, see \er{realJorA}. By Proposition \ref{KadA1} the latter induces an affine homeomorphism 
$\mathsf{K}: S(A)\raw S(A)$, which is automatically also uniformly continuous. Define 
\beq
\mathsf{W}= \mathsf{K}_{|P(A)}, \label{JKW}
\eeq
 and note that  $\mathsf{W}$
maps $P(A)$ to itself since $\mathsf{K}$ is a convex isomorphism and as such preserves the extreme boundary $\partial_e S(A)=P(A)$. Also,  $\mathsf{W}$ inherits uniform continuity from $\mathsf{K}$, cf.\ footnote \ref{wstar}. 
The fact that $\mathsf{W}$ preserves \tpies\ follows e.g.\ from \er{tpCstarA} and \er{JWA}, viz.
\begin{align}
\ta(\mathsf{W}(\om),\mathsf{W}(\om'))&= \inf\{\mathsf{K}(\om)(a)\mid a\in A, 0\leq a\leq 1_A,\mathsf{K}(\om')(a)=1\},\nn \\
&= \inf\{\om(\mathsf{J}(a))\mid a\in A, 0\leq a\leq 1_A,\om'(\mathsf{J}(a))=1\}\nn \\ &=
\inf\{\om(b)\mid b\in A, 0\leq b\leq 1_A,\om'(b)=1\} \nn \\ &= \ta(\om,\om'),
\end{align}
since if $a$ ranges over $0\leq a\leq 1_A$ then so does $b=\mathsf{J}(a)$. The same fact may also be proved from the expression \er{tpsh}, which in any case plays a crucial role later on, so that we first validate it:
\begin{lemma} \label{RT3.79} % Kitty Theorem 3.79
The expressions \er{tpCstarA} and \er{tpsh} for the \tp\ on $P(A)$ coincide, i.e.\ 
\begin{equation}
\ta(\om,\om')=\til{\om}(p(\om')), \label{tp3}
\end{equation}
for all $\om,\om'\in P(A)$,
where we abbreviate the carrier projection of $\om$ in $A^{**}$ by 
\begin{equation}
p(\om)\equiv \mathrm{carrier}(\om):=\inf\{e\in\CP(A^{**})\mid \til{\om}(e)=1\}. \label{defpom}
\end{equation}
 \end{lemma}
We first review some of the fine structure of the double dual $A^{**}$, identified as usual with the von Neumann algebra generated by the universal \rep\ (Pedersen 1979, \S\S 3.7--3.8):
\beq
A^{**}\cong (\oplus_{\rh\in S(A)}\pi_{\rh}(A))''\subset B(H_u)=B\left(\oplus_{\rh\in S(A)}H_{\rh}\right).
\label{Ped}
\eeq 
The following facts are well known and may be inferred from Theorem 5.19 in Alfsen \& Shultz (2001) and Lemmas 3.42  and 5.53 in Alfsen \& Shultz (2003), see also Takesaki (2002), \S III.2. A detailed derivation may be found in the proofs of Corollaries 4.32 and 4.33 in Rang (2019). 
\begin{enumerate}
\item Let  $\Om_{\om}\in H_{\om}$ be
 the cyclic vector from the GNS-construction $\pi_{\om}:A\raw B(H_{\om})$ induced by $\om\in P(A)$, seen as a vector in the universal \Hs\ $H_u$, i.e.\ $H_{\om}\subset H_u=\oplus_{\rh\in S(A)}H_{\rh}$. Then the carrier projection \er{defpom} is given by the one-dimensional projection on $\C\cdot\Om_{\om}$, i.e.
 \begin{equation}
p(\om)=|\Om_{\om}\ra\la\Om_{\om}|. \label{pom}
\end{equation}
  \item There exists a central projection $z$ in $A^{**}$ such that $zA^{**}$ is the \emph{atomic part} of $A^{**}$,\footnote{
The \emph{atomic part} of a \vna\ $M$ is given by $zM$, where $z$ is the supremum over all atoms in $\CP(M)$, defined as elements $f\neq 0$ for which $0\neq e\leq f$  implies $e=f$; then $z$ is automatically a central projection. As its name suggests, $zM$ is \emph{atomic} in that every nonzero element of its projection lattice $\CP(zM)$ dominates an atom.
 } viz.
 \begin{equation}
  z= \oplus_{[\om]\in P(A)/\sim}\,  c(\om), \label{cz1}
\end{equation}
where 
\beq
c(\om)=\inf\{e\in \CP(Z(A^{**}))\mid \til{\om}(e)=1\}. \label{defcom}
\eeq
 This is the \emph{central carrier} of $\om$, to be distinguished from 
in \er{defpom}: in  \er{defcom} the infimum is taken over all \emph{central} projections in $A^{**}$, so $p(\om)\leq c(\om)$.
 There is a simple dichotomy:
 \begin{align}
 c(\om)=c(\om') & \mbox{ iff }   \om\sim\om; \label{di1}\\
  c(\om)\perp c(\om') & \mbox{ iff }   \om\nsim\om,\label{di2}
 \end{align}
 where $\om\sim\om'$ means that $\om$ and $\om'$ are equivalent in that the corresponding {\sc gns}-\rep s $\pi_{\om}$ and $\pi_{\om'}$ are unitarily equivalent, and $\om\nsim\om$ means that this is not the case, see also \S\ref{GTP}. 
Eq.\ \er{di1} implies that the projection \er{cz1} is well defined, as is the ensuing decomposition 
\begin{equation}
 zA^{**}= \oplus_{[\om]\in P(A)/\sim}\,  c(\om) A^{**}. \label{cz2}
\end{equation}
\item Removing multiplicity, we  also have a (non-spatial) isomorphism of \vna s
 \begin{align}
  c(\om) A^{**}&\cong B(H_{\om}). \label{com}
 \end{align}
\item Each state $\rho\in S(A)$ extends uniquely to a normal state $\til{\rh}$ on $A^{**}$, seen as a \vna\ as in \er{Ped}, 
so that  $S(A)\cong S_n(A^{**})$. Restricting this  to pure states gives identifications
 \beq
 P(A)\cong P_n(A^{**})\cong P_n(zA^{**}), \label{PPn}
 \eeq
 as the non-atomic part $(1-z)A^{**}$ lacks normal pure  states. Combined with \er{com} this yields
\begin{equation}
[\om]\cong P_n(B(H_{\om})),\label{omPn}
\end{equation}
where $[\om]\subset P(A)$; this also gives a fresh perspective on Definition \ref{tpCAnorm} and preceding text.
\end{enumerate}
\emph{Proof of  Lemma \ref{RT3.79}}. This now immediately follows from 
Theorem 5.19 in Alfsen \& Shultz (2001), which we paraphrased around \er{oma} in \S\ref{GTP}, as well as from
eqs.\ \er{pom}, \er{di1}, and \er{di2}. \hfill $\Box$
\smallskip

As promised, we now also reprove the fact that, given a Jordan symmetry $\J$ and the ensuing Kadison symmetry $\K$, the map $\W$ defined by \er{JKW} preserves \tpies.

First, $\mathsf{J}:A\raw A$ extends to a normal Jordan automorphism $\J^{**}:A^{**}\raw A^{**}$, cf.\ Alfsen \& Shultz (2003), Proposition 2.74; here and in what follows, we tacitly extend \er{realJorA} to its complexification $J_{\C}:A\raw A$ and call this extension $J$ as well for simplicity. Since $\mathsf{J}^{**}$ is a Jordan map, it preserves squares and self-adjointness, and hence projections, as does its inverse. By \er{JWA} we have 
\beq
\widetilde{\W(\om')}((\J^{**})\inv(p(\om')))=\til{\om}'(\J^{**}((\J^{**})\inv(p(\om'))))=\til{\om}'(p(\om'))=1,
\eeq
so that $p_{\W(\om')}\leq ((\J^{**})\inv (p(\om'))$ by definition of the carrier projection $p(\om)$, see \S\ref{GTP}. Since Jordan maps also preserve positivity, this gives $\J^{**}(p_{\W(\om')})\leq p(\om')$. A similar argument gives 
\beq
\til{\om}'(\J(p_{\W(\om')}))=1,
\eeq
 and hence $p(\om')\leq \J^{**}(p_{\W(\om')})$,  again  by definition of the carrier projection, so that 
 \beq
 \J^{**}(p_{\W(\om')})= p(\om').
 \eeq
 From this and \er{tp3},  we compute
\begin{equation}
\ta(\W(\om),\W(\om'))=\widetilde{\W(\om)}(p_{\W(\om')})=\til{\om}(\J(p_{\W(\om')}))=\til{\om}(p_{\W(\om')})=\ta(\om,\om').
\end{equation}

We are now in a position prove the converse (and considerably more difficult) direction of Theorem \ref{Kitty}, i.e.\ from Wigner to Jordan. Suppose we have a Wigner symmetry \er{defWA}. 
\begin{lemma}\label{L22}
The map $\W$ preserves equivalence classes of pure states.
\end{lemma}
\noindent\emph{Proof.} 
This is a consequence of the assumption that  $\W$ preserves \tpies.
Indeed, it follows from Definition \ref{tpCAnorm} and preceding text that $\om\sim\om'$ iff there exists $\om''\in P(A)$ such that $\ta(\om,\om'')>0$ as well as  $\ta(\om',\om'')>0$: if $\ta(\om,\om')>0$ (whence $\om\sim\om')$ take $\om''=\om'$ and if $\ta(\om,\om)=0$ yet $\om\sim\om'$ take $\om''(a)=\la \Psi, \pi(a)\Psi\ra$ to be the vector state defined by (for example)\beq\Psi=(\Om_{\om}+\Om_{\om'})/\sqrt{2}.
\eeq 
If $\om\nsim\om'$, the the existence of such $\om''$ would lead to a contradiction because of the transitivity of $\sim$ (which is clear from its definition). This fact easily proves the lemma. \hfill $\Box$
\smallskip

\noindent
It follows from this lemma that $\W$ bijectively maps each equivalence class $[\om]$ to $[\W(\om)]$, preserving \tpies, and hence, by \er{omPn}, gives \tp-preserving bijections
\beq
\W_{\om}: P_n(B(H_{\om}))\raw P_n(B(H_{\W(\om)})). \label{Womega}
\eeq
Now Corollary  \ref{EqWigJor} has the following variation, replacing $\CP_1(H)$ by $P_n(B(H))$:
  \begin{corollary}\label{EqWigJorbis}
 For any \Hs s $H_1, H_2$ (necessarily of the same dimension), each bijection
\beq
\W_{12}: P_n(B(H_1)) \raw P_n(B(H_2)) \label{W12}
\eeq
that preserves \tpies\
is induced by a Jordan isomorphism
\begin{equation}
\J_{21}: B(H_2)\raw B(H_1)
\end{equation}
such that for all  $\om\in P_n(B(H_1))$ and $a\in B(H_2)$ one has
\begin{equation}
\W(\om)(a)=\om(\J(a)).\label{JW2}
\end{equation}
\end{corollary}
 This follows from Corollary  \ref{EqWigJor} by taking any unitary $u:H_2\raw H_1$, which, realizing pure states as vector states, also maps $P_n(B(H_2))$ to $P_n(B(H_1))$. This 
 turns \er{W12}  into a map
 $\W_1:  P_n(B(H_1)) \raw P_n(B(H_1))$, to which,  noting that unitarity preserves \tpies, Corollary  \ref{EqWigJor} applies. 
This corollary implies that each bijection $\W_{\om}$ in \er{Womega} is induced by a Jordan isomorphism \beq
\J_{\om}: B(H_{\W(\om}))\raw B(H_{\om}),\eeq 
which by \er{com} induces a  Jordan symmetry 
\beq
\J^{**}_{\om}:   c(\W(\om)) A^{**}\raw   c(\om) A^{**}. \label{4.25}
\eeq
Using \er{cz2}, the direct sum of all operators  \er{4.25} then gives a Jordan symmetry of $zA^{**}$, i.e.
 \beq
\J_z^{**}=\oplus_{[\om]\in P(A)/\sim}\, \J^{**}_{\om}: \: zA^{**}\raw  zA^{**}. \label{stst}
\eeq
The final step is to turn \er{stst} to a Jordan symmetry $\J:A\raw A$, and this is where the assumption of uniform continuity of $\W$ comes in. Recall from basic functional analysis that the image $\hat{A}\subset A^{**}$ of $A$ (seen as a Banach space) under the canonical embedding $A\hookrightarrow A^{**}$, $a\mapsto \hat{a}$, $\hat{a}(\phv)=\phv(a)$,  is given by those elements of $A^{**}$ that are $w^*$-continuous on $A^*$. If $A$ is a \ca, $\hat{A}$ consists of those elements of $A^{**}$ that are ($w^*$) uniformly continuous on $S(A)$, see footnote \ref{wstar}. This of course implies that elements of $\hat{A}$ are  uniformly continuous on $P(A)$, but a converse statement in this direction is far from obvious. Fortunately, 
Theorem 17 in Shultz (1982) gives what we need: 
\begin{lemma}\label{S82}
Under the canonical embedding $A\hookrightarrow A^{**}$ and hence $zA\hookrightarrow zA^{**}$, 
$z\hat{A}$ consists of  those elements of $zA^{**}$ for which the functions $\hat{a}$, $\widehat{a^*a}$, and $\widehat{aa^*}$ are  ($w^*$) uniformly continuous on $P(A)$. 
\end{lemma}
\newpage
\noindent Since Wigner symmetries \er{defWA} are uniformly continuous on $P(A)$, Lemma \ref{S82} implies that
$\J_z^{**}$ in \er{stst} maps $z\hat{A}$ to itself, and, being invertible, is a bijection. Thus
it only remains to be shown that 
\begin{align}
A &\stackrel{\cong}{\raw} z\hat{A};\label{52}\\
a&\mapsto z\hat{a} \label{53}
\end{align}
is an isomorphism of \ca s. Since $z$ is a central projection, the homomorphism property is obvious, as is injectivity. Surjectivity follows from \er{cz2}, \er{Ped},  and the fact that pure states separate points of $A$: for $a\neq b$ there is $\om\in P(A)$ such that $\om(a)\neq\om(b)$. It finally follows from \er{52} - \er{53} that $\J_z^{**}$, of which we already inferred that it is a Jordan symmetry of $z\hat{A}$, also defines a Jordan $\J$ symmetry of $A$. Keeping track of all isomorphisms, $\J$ implements $\W$ as in \er{JWA}.
\hfill $\Box$
\section{Orientations of (pure) state spaces}\label{shultz}
Theorem \ref{Shultz} is based on the concept of an \hi{orientation} of the pure state space of a \ca. This was introduced by Shultz (1982) by adaptation of the corresponding notion for the (total) state space of a \ca, which, in turn, was first defined by Alfsen, Hanche-Olsen, \& Shultz (1980), who acknowledge that the idea goes back to Connes (1974).  See also Alfsen \& Shultz (2001), Chapter 5, for a detailed treatment. Below we review the theory for pure and total state spaces in parallel and also provide some proofs omitted by Shultz (1982). We keep our standing assumption that \ca s are unital and refer to Rang (2019) for the non-unital case. 
 
The origin of the idea, and simultaneously its simplest case, lies in the \ca\ $M_2(\C)$ of complex $2\x 2$ matrices. We start with some  well-knows facts about the state space of $M_2(\C)$.\footnote{See, for example, Landsman (2017), \S\S 2.2 and 5.2.} Let $B^3=\{(x,y,z)\in\R^3\mid x^2+y^2+z^2\leq 1\}$ be the closed unit ball in $\R^3$, seen as a compact convex topological space, with extreme boundary $S^2=\{(x,y,z)\in\R^3\mid x^2+y^2+z^2= 1\}$.
\begin{proposition}\label{P3.1}
\begin{enumerate}
\item The state space $S(M_2(\C))$ of $M_2(\C)$ is affinely homeomorphic to $B^3$ 
through
\begin{align}
\rh: B^3 &\stackrel{\cong}{\raw} \mathcal{D}(\C^2);\label{B31}\\
(x,y,z)&\mapsto \half
\left(
\begin{array}{cc}
  1+ z & x-iy     \\
 x+iy & 1-z        
\end{array}
\right), \label{nicely1}
\end{align}
combined with the usual identification $\mathcal{D}(\C^2)\cong S(M_2(\C))$ via $\rh\lraw\om$, $\om(a)=\Tr(\rh a)$.
 \item Any affine bijection of $B^3\subset\R^3$  is given by restriction of some orthogonal  linear map $R\in O(3)$ (i.e.\ 
 $R\in GL_3(\R)$ with $R^T=R\inv$), and hence is necessarily a homeomorphism.
\item 
 On \er{B31} - \er{nicely1}, the extreme boundary $S^2\subset B^3$  is mapped to the pure state space $\CP_1(\C^2)\cong P(M_2(\C))$, and the \tpies\ \er{UtrUtr} on $\CP_1(\C^2)$ correspond to the function
 \begin{align}
 \ta:S^2\x S^2&\raw [0,1];\label{simonid1}\\
 ( \vec{x},\vec{y})&\mapsto \half(1+\la \vec{x},\vec{y}\ra)= \cos^2(\half\theta(\vec{x},\vec{y})). \label{simonid2}
 \end{align}
 where $\la \vec{x},\vec{y}\ra$ is the inner product on $\R^3$ and 
  $\theta(\vec{x},\vec{y})$ is the arc distance  between $\vec{x}$ and $\vec{y}$. 
 \item Any bijection of $S^2$ that preserves the \tpies\ \er{simonid1} - \er{simonid2}  is given by restriction of an orthogonal  linear map $R\in O(3)$, and hence is necessarily a homeomorphism.
\end{enumerate}
\end{proposition}
Since the orthogonal group $O(3)$ has two components, viz.\  $O_{\pm}(3)=\{R\in O(3)\mid\det(R)=\pm 1\}$, of which 
$O_+(3)=SO(3)$ contains the identity, both the `Kadison symmetries`  of $B^3$ (i.e.\ the
affine bijections) and the `Wigner symmetries' of $S^2$ (i.e.\ the bijections preserving \tpies) fall into two disjoint classes:
those that are induced by $R\in O_{\pm}(3)$. Those in $O_+(3)$ are orientation-preserving,  whereas those  in $O_-(3)$ are orientation-reversing. In the context of Wigner's Theorem the former, transferred back to bijections of $S(M_2(\C))$ or $P(M_2(\C))$, are induced by unitary operators on $\C^2$, whereas the latter are induced by anti-unitary operators, which in turn correspond to isomorphisms  and anti-isomorphisms of $M_2(\C)$, respectively (see e.g.\ Landsman, 2017, \S5.2). 
 
The generalization of this idea to arbitrary \ca s is based on the following dichotomy.\footnote{A \hi{face} in a convex set $K$ is a convex subset $F\subset K$ that is closed under ``purification", i.e.\ if $\om\in F$ and $\om=t\rh+(1-t)\sg$ for some $\rh,\sg\in K$ and $t\in (0,1)$, then $\rh\in F$ and $\sg\in F$. For example, if $\om$ is pure and $\om=t\rh+(1-t)\sg$, then by definition of purity  $\rh=\sg=\om$, so $F=\{\om\}$ is  a face. Conversely, if $\{\om\}$ is a face, then $\om$ is pure. 

If $(X,\ta)$ is a \tp\ space and $S\subset X$, then $S^{\perp}=\{x\in X\mid \ta(x,y)=0\, \forall y\in S \}$, and $S^{\perp\perp}=(S^{\perp})^{\perp}$.}
\begin{proposition}\label{dic}
Let $A$ be a (unital) \ca\ and let $\om$ and $\om'$ be pure states on $A$ ($\om\neq\om'$).
\begin{enumerate}
\item If $\om\sim\om'$, then the  \hi{face} $F(\om,\om')$ generated by $\om$ and $\om'$ within $S(A)$ (which by definition equals the intersection of all faces containing $\om$ and $\om'$) is affinely homeomorphic to $B^3$. 

Furthermore, the \hi{\tp\ space} 
\beq
T(\om,\om'):=\{\om,\om'\}^{\perp\perp}
\eeq generated by $\om$ and $\om$ in $P(A)$ 
is isomorphic to $S^2$ with \tpies\ \er{simonid1} - \er{simonid2}.

\item If $\om\nsim\om'$, then 
\begin{align}
F(\om,\om')&=\{t\om+(1-t)\om'\mid t\in [0,1]\};\label{72}\\
T(\om,\om')&=\{\om,\om'\}, \label{73}
\end{align}
so that $F(\om,\om')$ is affinely homeomorphic to $[0,1]$ and $T(\om,\om')$
 is isomorphic (as a \tp\ space) to its boundary $\{0,1\}$ with $\ta(x,y)=\dl_{xy}$.
\end{enumerate}
Finally, in both cases we have
\begin{equation}
T(\om,\om')=\partial_e F(\om,\om')= F(\om,\om')\cap P(A). \label{74}
\end{equation}
\end{proposition}
 The claims about $S(A)$ are Theorem 5.36 in Alfsen \& Shultz (2001), with a very clear proof.
 The claims about $P(A)$ are stated without proof in Shultz (1982). The proof, however, seems nontrivial to us, except for the final equality in \er{74}: for any face in a convex set $K$ one has $\partial_e F=F\cap \partial_e K$. \smallskip
 
\noindent\emph{Proof.} We invoke the following general result (Alfsen \& Shultz, 2001, Theorem 3.35):\footnote{In this theorem, which is about \vna s $M$, we took  $M=A^{**}$ and used $S(A)\cong S_n(A^{**})$, as always.} 
\begin{proposition}\label{AS335}
For any \ca\ $A$ there is a lattice (and in particular an order) isomorphism
\begin{align}
 \mathcal{F}(S(A))&\lraw \CP(A^{**}); \\
F &\lraw f;\\
f&=\mathrm{carrier}(F):= \inf\{e\in\CP(A^{**}) \mid  \til{\om}(e)=1 \, \forall \om\in F\};\\
F&= \{\om\in S(A)\mid\til{\om}(f)=1\} \label{79}
\end{align}
between the (complete) lattice $\mathcal{F}(S(A))$ of norm-closed faces in $S(A)$,  partially ordered by inclusion, and the   (complete) projection  lattice  $\CP(A^{**})$, partially ordered by $e\leq f$ iff $e=ef$. Under this isomorphism,\footnote{Which is even an isomorphism of \emph{orthomodular} lattice, see Proposition I.3.6.3 in Landsman (1998).}
 pure states in $S(A)$ correspond to atomic (= minimal) projections in $\CP(A^{**})$.
\end{proposition}
To prove the first part of Proposition \ref{dic}, take $f\equiv f(\om,\om')$ to be 
the (two-dimensional) projection onto the  linear span of the unit vectors $\Om_{\om}$ and $\Om_{\om'}$ in $H_u$ that represent $\om$ and $\om'$, respectively, cf.\  \er{Ped} and \er{oma}. Since
$f\Om_{\om}=\Om_{\om}$ and $f\Om_{\om'}=\Om_{\om'}$, the vector states on $A^{**}$
defined by $\Om_{\om}$ and $\Om_{\om'}$, i.e.\ $\til{\om}$ and $\til{\om}'$, are really states on the corner algebra $fA^{**}f$
in that they vanish on the other three blocks   $(1-f)A^{**}f$,  $fA^{**}(1-f)$, and $(1-f)A^{**}(1-f)$ in $A^{**}$.
Consequently, $\rh\in \{\om,\om'\}^{\perp}$ iff $\rh$ is atomic and $\mathrm{carrier}(\rh)\leq 1-f$, and hence  $\sg\in \{\om,\om'\}^{\perp\perp}$ iff $\sg$ is atomic and $\mathrm{carrier}(\sg)\leq f$. Thus $\sg$ is the vector state defined by some unit vector in the  linear span of $\Om_{\om}$ and $\Om_{\om'}$, or, in other words,  the density operator corresponding to $\sg$ is a one-dimensional projection onto $fH_u$. 

 To complete the argument we invoke Lemma 5.43 in Alfsen \& Shultz (2001):
 \begin{lemma}\label{L3.4}
If $\om\sim\om'$ and $\om\neq\om'$, then, with the projection $f\in\CP(A^{**})$ as just defined:
\begin{enumerate}
\item The  face $F\subset S(A)$ corresponding to the projection $f$ in Proposition \ref{AS335} is $F(\om,\om')$;
\item The corner algebra  $fA^{**}f$ is isomorphic to $M_2(\C)$ as a \ca.
\end{enumerate}
\end{lemma}
This implies that $F(\om,\om')$ is the state space of $fA^{**}f$, and since we saw that  $\{\om,\om'\}^{\perp\perp}$ is the pure state space of $fA^{**}f$, both  the first claim and eq.\ \er{74} for the case of equivalent states  follow.\footnote{It is also possible to prove this case without Proposition \ref{AS335}, instead relying on Theorem 5.19 in  Alfsen \& Shultz (2001) as  used in \er{oma} and surrounding text. This alternative approach reduces the proof to a calculation within $B(H)$, but to achieve this reduction one needs various results on so-called \emph{split faces}. See Rang (2019), Proposition 4.27.}. 
\smallskip

\noindent Part 2 of Proposition \ref{dic} may also be proved from Proposition \ref{AS335}, but there is a lighter argument, which we give instead.
By \er{tpineqstates} we have $\ta(\om,\om')=0$, so that we may invoke eq.\ I.(2.59) in Landsman (1998), which states that for any (mutually) orthogonal subset $S\subset P(A)$ one has
\begin{equation}
S^{\perp\perp}=\{\rh\in P(A)\mid \sum_{\sg\in S} \ta(\rh,\sg)=1\}.
\end{equation}
Taking $S=\{\om,\om'\}$, one has $\rh\in \{\om,\om'\}^{\perp\perp}$ iff $\ta(\rh,\om)+\ta(\rh,\om')=1$. Since $\om\nsim\om'$, it follows that
$0<\ta(\rh,\om)<1$ implies $\ta(\rh,\om')=0$, and so  $\rh\in \{\om,\om'\}^{\perp\perp}$ iff either $\ta(\rh,\om)=1$ or $\ta(\rh,\om')=1$. But in $P(A)$ one has $\ta(\rh,\om)=1$ iff $\rh=\om$, see Theorem I.2.8.2 in Landsman (1998).
Given \er{72}, this   validates \er{74} also for the case of inequivalent pure states. \hfill $\Box$

\begin{corollary}\label{C34}
Let $\K: S(A)\raw S(A)$ and $\W: P(A)\raw P(A)$  be  Kadison and Wigner symmetries of $A$, respectively, with 
$\W=\K_{|P(A)}$ as in \er{JKW},
and let  $\om$ and $\om'$ be pure states on $A$ ($\om\neq\om'$). Then 
\begin{align}
\K(F(\om,\om'))&=F(\W(\om),\W(\om'));\label{81}\\
\W(T(\om,\om'))&=T(\W(\om),\W(\om')).\label{82}
\end{align}
\end{corollary}
\noindent\emph{Proof.} Eq.\ \er{81} is trivial, since $\W(\om)=\K(\om)$ and $\K$ is an affine bijection. Eq.\  \er{82} follows from
the  fact that if $X\subset P(A)$, then $\W(X^{\perp})=\W(X)^{\perp}$ and hence $\W(X^{\perp\perp})=\W(X)^{\perp\perp}$. Indeed, 
 \begin{align}
        \W(X)^\perp &= \{\omega \in P(A) \mid \tau(\omega,\rho) = 0 \; \forall \rho \in \W(X) \} \nonumber \\
        &= \{\omega \in P(A) \mid \tau(\W^{-1}(\rho),\W^{-1}(\omega)) = 0 \; \forall \rho \in \W(X) \} \nonumber \\
        &= \{\omega \in P(A) \mid \tau(\sigma, \W^{-1}(\omega))=0 \; \forall \sigma \in X \} \nonumber \\
        &= \{ \W(\nu) \mid \nu \in P(A),\; \tau(\sigma,\nu)=0 \; \forall \sigma \in X \} \nonumber \\
        &= \W(X^\perp).
    \end{align}
Finally, the last claim of Proposition \ref{dic} follows from Lemma \ref{L22} and Proposition \ref{AS335}.\hfill $\Box$
\smallskip

\noindent Corollary \ref{C34} implies that $\K$ maps \hi{3-balls} (i.e.\ 
faces in $S(A)$ of the kind $F(\om,\om')$, $\om\sim\om'\neq\om$)  to each other through affine bijections, and $\W$ maps \hi{2-balls} (i.e.\ 
 \tp\ spaces in $P(A)$ of the kind $T(\om,\om')$, $\om\sim\om'\neq\om$)  to each other through bijections preserving \tpies.
We now equip each  3-ball in $S(A)$ and each 2-sphere in $P(A)$ with a standard orientation.\footnote{\label{densityfootnote} In differential geometry a connected manifold $M$ is called  
\hi{orientable} if it has an atlas whose transition functions all have positive Jacobian. If so, the equivalence class of such an atlas is an \hi{orientation} of $M$. If the atlas consists of a single chart, as is the case with  $F(\om,\om')$ with $\om\sim\om'\neq\om$, for which any affine bijection $\phv: B^3\raw F(\om,\om')$ defines a chart,  the equivalence class of the chart  is an orientation, as in Alfsen, Hanche-Olsen, \& Shultz (1980) and Alfsen \& Shultz (2001).
Although $S^2$ cannot be described by a single chart, we keep the idea that an orientation of $T(\om,\om')$ is simply an equivalence class of bijections $\psi: S^2\raw T(\om,\om')$ preserving \tpies\ (Shultz, 1982).}

First, an  \hi{orientation} of a  3-ball  $F(\om,\om')$ in $S(A)$  is an  equivalence class of affine bijections 
\beq
\phv: B^3\raw F(\om,\om'), \label{phvB3}
\eeq
where we assume that  $\om\sim\om'\neq\om$, and say that $\phv_1\sim\phv_2$ iff $\phv_2\inv\circ\phv_1: B^3\raw B^3$ lies in $SO(3)$, see Proposition \ref{P3.1}.2.
The \hi{standard orientation} of $F(\om,\om')$ is then defined via Lemma \ref{L3.4}.2, as follows. If  $f= f(\om,\om')$ is the projection described above the lemma, 
pick some isomorphism 
\beq
\al_f: fA^{**}f\raw M_2(\C), \label{alf}
\eeq
 pull this back to a map $\al_f^*: S(M_2(\C))\raw S(fA^{**}f)$, and use the isomorphism $S(M_2(\C))\cong B^3$ 
of Proposition \ref{P3.1}.1 as well as the identification $S(fA^{**}f)\cong F(\om,\om')$
to rewrite this as a map 
\beq
\phv_f\equiv \til{\al}_f^*: B^3\raw F(\om,\om'). \label{alfb3}
\eeq
 This map is an affine bijection by construction, and the crucial fact is that its equivalence class--and hence the orientation of $F(\om,\om')$ it defines--is independent of the choice of $\al_f$ as long as this is an isomorphism of \ca s. Indeed, if we start from another isomorphism $\beta_f: fA^{**}f\raw M_2(\C)$, the map $\al_f\circ\beta_f\inv: M_2(\C)\raw M_2(\C)$ is an automorphism of $M_2(\C)$ as a \ca,\footnote{Unitary implementability of automorphisms of any $B(H)$ is standard, see e.g.\  Theorem \ref{step3} above. For $H=\C^2$ an even simpler argument  based on group theory (instead of functional analysis) is available (Landsman, 2017, \S 5.2). } so that
 \beq
 \al_f\circ\beta_f\inv (a)=uau^* \label{87}
 \eeq
  for some unitary $u:\C^2\raw\C^2$, for all $a\in M_2(\C)$. As explained e.g.\ in Landsman (2017), \S 5.2, the pull-back of an automorphism $a\mapsto uau^*$ to $S(M_2(\C))\cong B^3$  is implemented by a rotation $R$ in $SO(3)$, so that
  $\til{\al}_f^*=\til{\beta}_f^*\circ R$, and hence $\al_f$ and $\beta_f$ define the same orientation of $F(\om,\om')$.
  This endows each  3-ball $F(\om,\om')$ in $S(A)$ with an orientation, called the \hi{standard orientation}.\footnote{This also defines an orientation of $S(A)$ in a certain global sense defined in Alfsen \& Shultz, Chapter 5, but we will not need this. Briefly, the idea is that the set $\mathrm{Param}(S(A))$ of all maps $\phv: B^3\raw S(A)$ such that $\phv(B^3)$ is some  3-ball and $\phv$ as an affine bijection onto its image forms an $SO(3)$ bundle as well as an $O(3)$ bundle in the obvious way, giving rise to a $\Z_2$-bundle $\mathrm{Param}(S(A))/SO(3)\raw \mathrm{Param}(S(A))/O(3)$. A global orientation of $S(A)$ is a continuous cross-section of this bundle, and the procedure in the main text in fact canonically specifies such a  global orientation.} 
  
  A similar procedure  endows each 2-sphere $T(\om,\om')$ inside $P(A)$ with a standard orientation
  \beq
\psi_f\equiv  (\til{\al}_f^*)_{|S^2}: S^2\raw T(\om,\om'), \label{psif}
  \eeq   
 which is simply the restriction of the map \er{alfb3} to the (extreme) boundary $S^2=\partial_e B^3$ of $B^3$. 
 More precisely, for pure states the affine bijections \er{phvB3} are replaced by  bijections 
  \beq
  \psi: S^2\raw T(\om,\om') \label{89}
  \eeq
   preserving \tpies;  we define  $\ps_1\sim\ps_2$ iff $\ps_2\inv\circ\ps_1: S^2\raw S^2$ lies in $SO(3)$, cf.\  Proposition \ref{P3.1}.4,  and the standard orientation of $ T(\om,\om')$ is the equivalence class of \er{psif}.
\section{Proof of  Theorem \ref{Shultz} (Shultz)}
We are now in a position to understand and prove  Theorem \ref{Shultz}, due to Shultz (1982). Our proof, based on our own Theorem \ref{Kitty}, is quite different from Shultz's, though clearly inspired by it.  

Let
us first explicate the assumption that $\W$ (and similarly $\K$) preserves orientation. Define
\beq
 f'=f(\W(\om), \W(\om'))=(J^{**})\inv (f(\om,\om')), \label{fprime}
\eeq
where $\W:P(A)\raw P(A)$ is a Wigner symmetry,  $\om\sim\om'\neq\om$ are two equivalent states as before,
with associated projection $f=f(\om,\om')\in \CP(A^{**})$, and $\J^{**}: A^{**}\raw  A^{**}$ is the unique normal extension of the Jordan symmetry $\J$ associated to $\W$ by
 Theorem \ref{Kitty}. It follows from \er{79} and \er{JKW} that \er{fprime} is the correct expression for $f'$.
We then have the following  maps besides \er{psif}: 
\begin{align}
\psi_{f'}: S^2 &\raw T(\W(\om), \W(\om'));\\
\W_{\om,\om'}: T(\om,\om')&\raw  T(\W(\om), \W(\om')),
\end{align}
where, as in  \er{psif},
 the map $\psi_{f'}$ is induced by some isomorphism of \ca s 
 \beq
 \al_{f'}: f' A^{**}f'\raw M_2(\C),
 \eeq
and $\W_{\om,\om'}$ is just the restriction of $\W$ to $T(\om,\om')$, see \er{82}. Now recall  the end of \S\ref{shultz}.
\begin{definition}
A Wigner symmetry $\W$  \hi{preserves orientation} if for all pairs $\om\sim\om'\neq\om$ in $P(A)$ the maps $\W_{\om,\om'}\circ \psi_f$ and $\psi_{f'}$, both from $B^3$ to $T(\W(\om), \W(\om'))$, are equivalent, i.e.\
if the map
\beq
\psi_{f'}\inv\circ  \W_{\om,\om'}\circ \psi_f: S^2\raw S^2
\eeq
 is induced by some $R\in SO(3)$. Similarly, a
 Kadison symmetry $\K$ of $S(A)$ preserves orientation if
\beq
\phv_{f'}\inv\circ  \K_{\om,\om'}\circ \phv_f: B^3\raw B^3
\eeq is induced by some $R\in SO(3)$, again  for all pairs $\om\sim\om'\neq\om$,  cf.\  \er{81} and \er{alfb3}. 
 \end{definition}
 Note that although the assumption that $\W$ (or $\K$) be orientation-preserving (reversing) has now been well \emph{defined}, as explained after the proof of Corollary \ref{C34} neither of these alternatives need to \emph{hold}: in fact,  for typical \ca s most Jordan symmetries neither preserve nor reverse orientation, except for $A=B(H)$, where Theorem \ref{step3} shows that either one or the other must be the case. Thus Theorem \ref{Shultz} sharpens  Theorem \ref{Kitty}  under fairly restrictive extra assumptions.
\smallskip

\noindent\emph{Proof of  Theorem \ref{Shultz}.} We know from Theorem \ref{Kitty} that, irrespective of any orientability requirement, 
$\W$ comes from a Jordan symmetry $\J$, which, by Proposition \ref{KadA2}, in turn induces a Kadison symmetry $\K$.
Concerning the latter, one has the decisive Theorem 5.71 in Alfsen \& Shultz (2001), which for convenience we combine with Proposition \ref{KadA1} so as the obtain the following:
\begin{theorem}\label{Shultz2}
There is a bijective correspondence $\mathsf{K}\lraw \al$ between orient\-ation-preserving (reversing)  Kadison symmetries \er{defKA1} of $A$
and (anti-) automorphisms $\al$ of $A$, such that for each $\om\in S(A)$,
\begin{equation}
\mathsf{K}(\om)=\om\circ \al. \label{KJA3}
\end{equation}
\end{theorem} 
Since $\W$ and $\K$ are related by \er{JKW}, i.e.\ $\mathsf{W}= \mathsf{K}_{|P(A)}$, it follows from our very definition of the (standard) orientations of $P(A)$ and $S(A)$ that $\W$ and $\K$ both preserve or reverse orientation  (or both fail to do so), so that, given Theorem \ref{Kitty},  Theorem \ref{Shultz} is  equivalent to  Theorem \ref{Shultz2}. \hfill $\Box$
\smallskip

\noindent One may also prove  Theorem \ref{Shultz} directly from Theorem \ref{Kitty} by
adapting and rephrasing the proof of  Theorem 5.71 in Alfsen \& Shultz (2001), \emph{mutatis mutandis}. This looks  as follows. 

First, each of the maps \er{alfb3} and  \er{psif} is defined as the compositions of five isomorphisms:
\begin{align}
\phv_f:& & B^3\stackrel{\rh}{\longrightarrow}\mathcal{D}(\C^2)\stackrel{\footnotesize{\Tr}}{\longrightarrow}S(M_2(\C))\stackrel{\al_f^*}{\longrightarrow} S(fA^{**}f)  \stackrel{(f - f)}{\longrightarrow}
\til{F}(\om,\om')\stackrel{\sim}{\longrightarrow} F(\om,\om')  ;\label{6.8}\\
\psi_f:& & S^2\stackrel{\rh}{\longrightarrow}\CP_1(\C^2)\stackrel{\footnotesize{\Tr}}{\longrightarrow}P(M_2(\C)) \stackrel{\al_f^*}{\longrightarrow}
P(fA^{**}f) \stackrel{(f-f)}{\longrightarrow}
\til{T}(\om,\om')\stackrel{\sim}{\longrightarrow} T(\om,\om'), \label{6.9}
\end{align}
where in the first line $\rh$ is the parametrization \er{B31}, which in the second line is tacitly restricted to $S^2\subset B^3$ so as to land in $\CP_1(\C^2)\subset \mathcal{D}(\C^2)$, and the $\Tr$ in both lines is the usual identification of density matrics and (normal) states, i.e.,
$\rh(\vec{x})\mapsto (\om: a\mapsto\Tr(\rh(\vec{x}) a))$. Furthermore, we have
\begin{align}
\til{T}(\om,\om')&\subset \til{F}(\om,\om')\subset S_n(A^{**})\cong S(A); \label{95}\\
T(\om,\om')&\subset F(\om,\om')\subset S(A). \label{96}
\end{align}
As in item 4 containing \er{PPn} we use the isomorphisms $S(A)\cong S_n(A^{**})$ and its restriction 
to pure states $P(A)\cong P_n(A^{**})$, written as $\rh\mapsto \til{\rh}$ and hence denoted by $\sim$ in \er{6.8} - \er{6.9}, as well as
\begin{align}
\til{F}(\om,\om')&= \{\til{\rh}\in S_n(A^{**})\mid\til{\rh}(f)=1\};\\
F(\om,\om')&= \{\rh\in S(A)\mid\til{\rh}(f)=1\},
\end{align}
cf.\ \er{79},
where $f$ is the projection in $A^{**}$ defined above Lemma \ref{L3.4}. Similarly for pure states:
\begin{align}
\til{T}(\om,\om')&= \{\til{\rh}\in P_n(A^{**})\mid\til{\rh}(f)=1\};\\
T(\om,\om')&= \{\rh\in P(A)\mid\til{\rh}(f)=1\}.
\end{align}
In this context, the isomorphism $(f-f)$ in  \er{6.8}, and by restriction to pure states also the one in \er{6.9}, is given by 
identifying $\til{\rh}\in \til{F}(\om,\om')$ with $\hat{\rh}\in S(fA^{**}f)$ via $\til{\rh}(a)=\hat{\rh}(faf)$, where $a\in A^{**}$. Note that 
$f$ is the unit in $fA^{**}f$ and that, since $\al_f$ is an isomorphism, $ S(fA^{**}f)= S_n(fA^{**}f)$. 
In the situation of Theorem \ref{Kitty}, where $\K=\J^*$ and $\W=\K_{|P(A)}$, the following diagram commutes:
\beq
\begin{tikzcd}
P(fA^{**}f) \arrow[d, "\J_f^*"']\arrow[r, "f-f"] & \til{T}(\om,\om') \arrow[r, "\sim"] & T(\om,\om') \arrow[d, "\W_{\om,\om'}"]\\ 
P('fA^{**}f') \arrow[r, "f-f"] &\til{T}(\W(\om),\W(\om')) \arrow[r, "\sim"] & T(\K(\om),\K(\om')),
\end{tikzcd}
\eeq
where $\J^*_f$ is the restriction of $\W:P(A)\raw P(A)$, seen as a map $\til{\W}: P_n(A^{**})\raw P_n(A^{**})$, to
$P(fA^{**}f)$.
Consequently, in terms of the isomorphism $\til{\psi}_f=\al_f^*\circ \Tr\circ\rh: S^2\raw P(fA^{**}f)$, the map $\W$ preserves orientation if all maps $\psi_{f'}\inv\circ  \W_{\om,\om'}\circ \psi_f: S^2\raw S^2$ preserve orientation, i.e., if 
in the diagram
\beq
\begin{tikzcd}
S^2\arrow[d, "R"'] \arrow[r, " \til{\psi}_f"] & P(fA^{**}f) \arrow[d, "\J_f^*"] \\
S^2 \arrow[r, " \til{\psi}_{f'}"] 
&  P(f'A^{**}f')
\end{tikzcd} \label{105}
\eeq
the downward arrow $R$ on the left preserves orientation. This is settled by the following lemma:
\begin{lemma}
Let $B$ and $C$ be \ca s isomorphic to $M_2(\C)$, with specific isomorphisms 
\begin{align}
\beta:B\raw M_2(\C); &&\gm: C\raw M_2(\C),
\end{align} and let $\J:C\raw B$ be a Jordan map with pullback $\J^*: P(B)\raw P(C)$.  Also, define 
\begin{align}
\til{\beta}^*=\beta^*\circ\Tr\circ\rh:S^2\raw P(B); && \til{\gm}^*=\gm^*\circ\Tr\circ\rh:S^2\raw P(C).
\end{align} 
Then $\J$ is an (anti-) isomorphism of \ca s iff the map $R$ in the commutative diagram
\beq
\begin{tikzcd}
S^2\arrow[d, "R"'] \arrow[r, "\til{\beta}^*"] & P(B) \arrow[d, "\J^*"] \\
S^2 \arrow[r, " \til{\gm}^*"] 
&  P(C).
\end{tikzcd}
\eeq
preserves (reverses) orientation, that is, if $R\in SO(3)=O_+(3)$ ($R\in O_-(3)$). 
\end{lemma}
The proof is almost literally the argument surrounding \er{87}. Returning to the setting of Theorem \ref{Kitty}, it follows that
the Jordan symmetry $\J: A\raw A$ is an  (anti-) isomorphism of \ca s iff the downward arrows  on the left  in all diagrams \er{105}, where $f$ and $f'$ are defined in terms of $\om\sim\om'\neq \om$, running through $P(A)$, and $\J$ as before, 
preserve (reverse) orientation. Hence if $\J$ is an   (anti-) isomorphism of \ca s, then $\W$ by definition
preserves (reverses) orientation. 

For the converse implication, take a single pure state $\om\in P(A)$ and the associated face
 \beq
 F(\om)=S_n(B(H_{\om}))\subset S_n(A^{**})\cong S(A).
 \eeq
 This face contains all 3-balls $F(\om,\om')$ for  $\om\sim\om'\neq \om$, and hence if $\W$ preservers (reverses) orientation, then so do all maps $\W_{\om,\om'}$. Furthermore, by \er{defcom} and \er{JWA}, i.e.\ $\W=\J^*$,  we have
 \begin{align}
c(\W(\om))& =\inf\{e\in \CP(Z(A^{**}))\mid \til{\om}(\J^{**}(e))=1\}\nn \\
&= \inf\{(\J^{**})\inv e\in \CP(Z(A^{**}))\mid \til{\om}(e)=1\}\nn \\ & =(\J^{**})\inv \inf\{( e\in \CP(Z(A^{**}))\mid \til{\om}(e)=1\}\nn \\
&= (\J^{**})\inv c(\om),
\end{align}
since  $\J^{**}$ and  $(\J^{**})\inv$ are normal and hence preserve suprema and infima. 
Therefore, 
\beq
\J^{**}: c(\om)A^{**}\raw c(\W\inv(\om))A^{**}, \label{Jcomcom} 
\eeq
is an isomorphism of Jordan algebras.   By \er{com}, eq.\ \er{Jcomcom} is equivalent to a Jordan isomorphism 
\beq
\til{J}_{\om}: B(H_{\om})\raw B(H_{\W\inv(\om)}).
\eeq
Theorem \ref{step3} implies that this map is  an isomorphism or an anti-isomorphism or both (which is the case if $\dim(H_{\om})=1$, in which case there are no $\om'$ such that  $\om\sim\om'\neq \om$). 
If $\W$ preserves orientation, then by the first part of the proof $\til{J}_{\om}$ cannot be an anti-isomorphism (for in that case $\W$ would reverse the orientation of all $T(\om,\om')$),
and hence it must be an isomorphism. This is true for all $\om\in P(A)$, so that each $\til{J}_{\om}$ is an isomorphism of \ca s, which implies that the restriction of  $\J^{**}$ to each subspace $c(\om)A^{**}$ of $A^{**}$ is an isomorphism onto its image 
 $c(\W\inv(\om))A^{**}$. It then follows from \er{cz2} that  $\J_z^{**}:   zA^{**} \raw zA^{**}$ 
 is an  isomorphism of \ca s, and as in the last part of the proof of Theorem \ref{Kitty} this implies that $\J:A\raw A$ is an isomorphism. Similarly, if $\W$ reverses orientation, then $\J$ is an anti-isomorphism of \ca s.\hfill $\Box$

\section{Proof of Lemma \ref{Thomsen} (Thomsen)}
 \label{herstein}
 This lemma is the main result in Thomsen (1982), whose paper stands in a long tradition going back to Jacobson \& Rickart (1950), with important contributions by Kadison (1951) and Herstein (1956). See also  Bratteli \& Robinson (1987), Proposition 3.2.2 for a summary.
 Our aim is to make all steps explicit and thus provide a complete proof of this crucial result. 
 Throughout this section $A\subset B(H)$ is a unital  \ca\ and $\J_{\C}\equiv \J:A\raw A$ is a Jordan symmetry,\footnote{See Rang (2019), \S 2.4 for the non-unital case. Also, recall that for us a homomorphism is a $\mbox{}^*$-homomorphism.}  although for many results it actually suffices that $\J$ is merely a homomorphism and $A$ may often  be just a  *-algebra; these (irrelevant) generalizations are  clear from the proofs. 
 We prove the lemma by explicitly constructing the projections $p_1$, $p_2$ and $p_3$. To do this, we need to do some preparatory work in the form of Lemma \ref{lemma:jordan_equalities} and Lemma \ref{lemma:A_1*A_2=0}. We use the following notation. We define for $a,b\in A$:
    \begin{align}
        a^b = i[\J(ab)-\J(a)\J(b)]; && a_b &= i[\J(ab)-\J(b)\J(a)].
    \end{align}
    It is easy to see that if $a,b\in {A}\sa$, then $a^b,a_b\in {B}\sa$. Indeed (and similarly for $a_b$):
    \begin{align}
        (a^b)^* &= \big(i[\J(ab)-\J(a)\J(b)]\big)^* 
        = -i[\J(b^*a^*)-\J(b^*)\J(a^*)] \nonumber \\
        &= -i[\J(ba+ab)-\J(ab)-\J(b)\J(a)] 
        = -i[\J(b)\J(a)+\J(a)\J(b)-\J(ab)-\J(b)\J(a)] \nonumber \\
        &= i[\J(ab)-\J(a)\J(b)] =a^b.
    \end{align}
Using elementary linear algebra, we can deduce several useful relations concerning $a^b$ and $a_b$. These relations will be useful in the proof of Lemma \ref{lemma:A_1*A_2=0}. Lemma \ref{lemma:jordan_equalities} is due to Herstein (1956).

\begin{lemma}\label{lemma:jordan_equalities}
For all $a,b,c\in A$:
    \begin{align*}
& (i)\quad \J(aba)=\J(a)\J(b)\J(a);  &
& (ii)\quad  \J(abc+cba)=\J(a)\J(b)\J(c)+\J(c)\J(b)\J(a);\\
& (iii) \quad a^ba_b=0; &
     & (iv)\quad   a^b\J(c)a^b=ia^b\J((ab-ba)c);\\
& (v) \quad a^b\J((ab-ba)c)a_b=0; &
& (vi) \quad a^b\J(ab-ba)\J(c)\J(ab-ba)a_b=0; \\
&  (vii) \quad  a^b+a^c=a^{b+c};  &&   (viii) \quad a_b+a_c=a_{b+c};\\
&  (ix) \quad a^c+b^c=(a+b)^c; &&  (x) \quad a_c+b_c=(a+b)_c.
    \end{align*}
\end{lemma}
\emph{Proof.}
    \begin{enumerate}[label=(\roman*)]
        \item For all $a,b\in A$ we have
   $aba=2(a\circ(a\circ b))-b\circ(a \circ a)$,
        which shows that
            \begin{align}\label{eqn:jordan_aba}
                \J(aba) &= 2\J\big((a\circ(a\circ b))\big)-\J\big(b\circ(a \circ a)\big) \nonumber \\
                &= 2\big(\J(a)\circ(\J(a)\circ\J( b))\big)-\J(b)\circ\big(\J(a) \circ \J(a)\big) \nonumber \\
                &= \J(a)\J(b)\J(a).
            \end{align}
        \item Because for all $a,b,c\in A$ we have
                $abc + cba = (a+c)b(a+c) - aba - cbc$,
   eq.\ \eqref{eqn:jordan_aba} yields
            \begin{align}\label{eqn:jordan_(abc+cba)}
                \J(abc + cba) &= \J\big((a+c)b(a+c)\big) - \J(aba) - \J(cbc) \nonumber \\
                &= \J(a+c)\J(b)\J(a+c) - \J(a)\J(b)\J(a) - \J(c)\J(b)\J(c) \nonumber \\
                &= \J(a)\J(b)\J(c) + \J(c)\J(b)\J(a).
            \end{align}
        \item It now follows from \eqref{eqn:jordan_aba} and \eqref{eqn:jordan_(abc+cba)} that $a^ba_b=0$ for all $a,b\in A$:
            \begin{align}\label{eqn:a^ba_b=0}
                a^ba_b &= -\big(\J(ab)-\J(a)\J(b)\big)\big(\J(ab)-\J(b)\J(a)\big) \nonumber \\
                &= -\J(ab)^2 + \big(\J(ab)\J(b)\J(a) + \J(a)\J(b)\J(ab)\big) - \J(a)\J(b)^2\J(a) \nonumber \\
                &= -\J(ab)\circ\J(ab) + \J(ab^2a+abab) - \J(ab^2a) \nonumber \\
                &= -\J(abab) + \J(ab^2a+abab) - \J(ab^2a) = 0.
            \end{align}
        \item We can now use \eqref{eqn:jordan_(abc+cba)} and the Jordan property to see that
        \begin{align}
            &\J(c)a^b = i\J(c)\J(ab)-i\J(c)\J(a)\J(b) \nonumber \\
            &\quad= i\J(c)\J(ab)+i\J(b)\J(a)\J(c)-i\J(cab+bac) \nonumber \\
            &\quad= i\J(c)\J(ab)+i\J(b)\J(a)\J(c)-i\J\big(c(ab)+(ab)c\big)+i\J\big((ab-ba)c\big) \nonumber \\
            &\quad= i\J(c)\J(ab)+i\J(b)\J(a)\J(c)-i\J(c)\J(ab)-i\J(ab)\J(c)+i\J\big((ab-ba)c\big) \nonumber \\
            &\quad= i\big(\J(ab)-\J(a)\J(b)\big)\J(c)+i\J\big((ab-ba)c\big) \nonumber \\
            &\quad= a_b\J(c)+i\J((ab-ba)c).
        \end{align}
        Now multiply both sides of the equation from the left with $a^b$ and use \eqref{eqn:a^ba_b=0} to conclude that
        \begin{align}\label{eqn:a^bphi(c)a^b}
            a^b\J(c)a^b &= a^ba_b\J(c)+ia^b\J\big((ab-ba)c\big)
            = ia^b\J\big((ab-ba)c\big).
        \end{align}
        \item Multiply equation \eqref{eqn:a^bphi(c)a^b} from the right with $a_b$ and use equation \eqref{eqn:a^ba_b=0} to conclude that
        \begin{equation}\label{eqn:a^bphi((ab-ba)c)a^b}
            a^b\J\big((ab-ba)c\big)a_b = -i a^b\J(c)a^ba_b = 0.
        \end{equation}
        \item Use equation \eqref{eqn:jordan_aba} and \eqref{eqn:a^bphi((ab-ba)c)a^b} and replace $c$ with $c(ab-ba)$ to conclude that\begin{equation}
            0 = a^b\J\big((ab-ba)c(ab-ba)\big)a_b = a^b\J(ab-ba)\J(c)\J(ab-ba)a_b.
        \end{equation}
        \item[(vii)] This follows immediately from the linearity of $\J$, as do (viii) - (x). \hfill $\Box$
    \end{enumerate}
\begin{lemma}\label{lemma:A_1*A_2=0}
For all   quadruples $a,b,c,d\in A$ one has
    \begin{equation}
        \big(\J(ab)-\J(a)\J(b)\big)\cdot \big(\J(cd)-\J(d)\J(c)\big)=0, \label{K1}
    \end{equation}
\end{lemma}
\emph{Proof.}  Let $P(A'')$ be the set of pure states on $A''$. Since the atomic representation
$\pi_a=\oplus_{\omega\in P(A'')}\pi_\omega$ is faithful (or, equivalently, because pure states separate points of $A''$),
eq.\ \er{K1} holds iff 
        \begin{equation}
            \big((\pi_\omega\circ\J)(ab)-(\pi_\omega\circ\J)(a)(\pi_\omega\circ\J)(b)\big)\cdot 
            \big((\pi_\omega\circ\J)(cd)-(\pi_\omega\circ\J)(d)(\pi_\omega\circ\J)(c)\big) = 0,
        \end{equation}
        for all $a,b,c,d\in A$ and $\omega \in P$. Since $\pi_{\om}$ is irreducible, this reduces the proof of \er{K1} to the case where $A''\subset B(H)$ is a factor. Since $\J(A)=A$, Lemma \ref{lemma:jordan_equalities} part (vi) implies that
    \begin{equation}
        a^b\J(ab-ba)\cdot A''\cdot\J(ab-ba)a_b=0.
    \end{equation}
   Now in any factor $M$ the condition $aMb=0$ implies that $a=0$ or $b=0$,\footnote{See e.g.\ Corollary 3 of Part I, Chapter 1, \S 4 on page 7 of Dixmier (1981).} so that
for all $a,b\in A$,
    \begin{equation}\label{eqn:a^bf(ab-ba)=0}
        a^b\J(ab-ba)=0 \text{ or }\J(ab-ba)a_b=0.
    \end{equation}
    We may use the Jordan property of $\J$ to rewrite
    \begin{align}\label{eqn:phi(ab-ba)=-i(a^b+a_b)}
        \J(ab-ba) = 2\J(ab) - \J(ab+ba)
        = 2\J(ab) - \J(a)\J(b) - \J(b)\J(a) 
        =-i(a^b+a_b).
    \end{align}
    First assume that $a^b\J(ab-ba)=0$ in  \eqref{eqn:a^bf(ab-ba)=0}. Using \eqref{eqn:phi(ab-ba)=-i(a^b+a_b)} and  \eqref{eqn:a^ba_b=0}, we find
    \begin{equation}
        0 = a^b\J(ab-ba) = -ia^b(a^b+a_b) = -i(a^b)^2,
    \end{equation}
    hence $(a^b)^2=0$.
    Similarly, the other possibility in  \eqref{eqn:a^bf(ab-ba)=0} implies $(a_b)^2=0$. If $a,b\in{A\sa}$, we know that $a^b$ and $a_b$ are self-adjoint. In that case $(a^b)^2=|a^b|^2=0$ or $(a_b)^2=\abs{a_b}^2=0$. Therefore, we can conclude that for all $a,b\in{A\sa}$:
    \begin{equation}\label{eqn:a^b=0_or_a_b=0}
        a^b=0 \text{ or } a_b=0.
    \end{equation}
    This shows that for all $a,b\in{A\sa}$ and $c\in A$:
    \begin{equation}
        a_b\J(c)a^b=0.
    \end{equation}
    Using this relation and Lemma \ref{lemma:jordan_equalities} parts (vii) - (x) we see that for all $a,b,d\in{A\sa}$ and $c\in A$:
    \begin{align}
        0 & =a_{b+d}\J(c)a^{b+d} = a_b\J(c)a^b + a_b\J(c)a^d + a_d\J(c)a^b + a_d\J(c)a^d = a_b\J(c)a^d + a_d\J(c)a^b.
    \end{align}
    Combining this with \eqref{eqn:a^b=0_or_a_b=0} implies that for all $a,b,d\in{A\sa}$ we have $a_d=0$ or $a^b=0$, and hence
    \begin{equation}
        a_da^b=0.
    \end{equation}
    Replacing $a$ by $a+c$ and again using Lemma \ref{lemma:jordan_equalities} parts (vii) - (x)  we find that for all $a,b,c,d\in{A\sa}$:
    \begin{align}\label{eqn:a_dc^b+c_da^b=0}
        0 &= (a+c)_d+(a+c)^b = a_da^b + a_dc^b + c_da^b + c_dc^b = a_dc^b + c_da^b.
    \end{align}
    We know that $a_d=0$ or $a^b=0$, and $c^b=0$ or $c_d=0$, hence \eqref{eqn:a_dc^b+c_da^b=0} implies that $a^b=0$ or $c_b=0$. So for all $a,b,c,d\in{A\sa}$ we have
    \begin{equation}
        a^bc_d=0.
    \end{equation}
    Now let $a,b,c,d\in A$ and write these as $a=a_1+ia_2,\dots,d=d_1+id_2$ with $a_1,a_2,\dots,d_2,d_2\in {A\sa}$. It then follows from Lemma \ref{lemma:jordan_equalities}  parts (vii) - (x)  and the previous relation that
    \begin{align*}
        &\big(\J(ab-\J(a)\J(b)\big)\cdot \big(\J(cd)-\J(d)\J(c)\big)  
   = -a^bc_d  = (a_1+ia_2)^{b_1+ib_2}(c_1+ic_2)_{d_1+id_2}  \\
        &\quad = (a_1^{b_1}+ia_1^{b_2}+ia_2^{b_1}-a_2^{b_2})\cdot ({c_1}_{d_1}+i{c_1}_{d_2}+i{c_2}_{d_1}-{c_2}_{d_2})  = 0. \tag*{$\Box$}
    \end{align*}
\begin{corollary}\label{cor:a_bc^d=0}
    For all quadruples $a,b,c,d\in A$ we have
    \begin{equation}
        \big(\J(ab)-\J(b)\J(a)\big)\cdot \big(\J(cd)-\J(c)\J(d)\big)=0.
    \end{equation}
\end{corollary}

\emph{Proof.}
From Lemma \ref{lemma:A_1*A_2=0}:
    \begin{multline}
        \big(\J(ab)-\J(b)\J(a)\big)\cdot \big(\J(cd)-\J(c)\J(d)\big) \\
        = \Big(\big(\J(d^*c^*)-\J(d^*)\J(c^*)\big)\cdot \big(\J(b^*a^*)-\J(a^*)\J(b^*)\big) \Big)^* =0.
    \end{multline}
\hfill $\Box$

\noindent Using Lemma \ref{lemma:A_1*A_2=0} and Corollary \ref{cor:a_bc^d=0}, we are finally able to prove Lemma \ref{Thomsen}.
    Define
    \begin{align}
        A_1 &= \{\J(ab)-\J(a)\J(b) \mid a,b \in A \}\subset A, \\
        A_2 &= \{\J(ab)-\J(b)\J(a) \mid a,b \in A \}\subset A.
    \end{align}
    Also define the projections
    \begin{equation}
        q_1 = \Big[\bigcap_{\alpha \in A_1}\ker{\alpha}\Big], \quad q_2 = \Big[\bigcap_{\alpha \in A_2}\ker{\alpha}\Big], \quad q_3 = \Big[\bigcap_{\alpha \in A_1\cup A_2}\ker{\alpha}\Big], \label{qqq}
    \end{equation}
    where, as usual,  $[\dots]$ denotes the orthogonal projection onto the closed subspace of $B(H)$ between the brackets. Now define (not confusing the prime with the sign for the commutant):
    \begin{align}
        A_1' &=\{\J(ab)-\J(a)\J(b) \mid a,b \in(A_1)\sa \};\\
        A_2' &=\{\J(ab)-\J(a)\J(b) \mid a,b \in(A_2)\sa \},
    \end{align}
    as well as the ensuing projections, quite analogously to \er{qqq},
    \begin{equation}
        q_1' = \Big[\bigcap_{\alpha \in A_1'}\ker{\alpha}\Big], \quad q_2' = \Big[\bigcap_{\alpha \in A_2'}\ker{\alpha}\Big], \quad q_3' = \Big[\bigcap_{\alpha \in A_1'\cup A_2'}\ker{\alpha}\Big].
    \end{equation}
    \emph{Claim:}
        We have $q_1=q_1'$, $q_2=q_2'$ and $q_3=q_3'$.\smallskip
        
  \noindent  \emph{Proof of claim:}
        We prove that $q_1=q_1'$; the other two identities are proven similarly. First note that $A_1'\subset A_1$, and hence
        \begin{equation}
            \bigcap_{\alpha \in A_1}\ker{\alpha}\subseteq   \bigcap_{\alpha\in A_1'}\ker{\alpha}.
        \end{equation}
       This shows that $q_1 \leq q_1'$, so it only remains to be shown that
        \begin{equation}
            \bigcap_{\alpha\in A_1'}\ker{\alpha} \subseteq \bigcap_{\alpha \in A_1}\ker{\alpha}.
        \end{equation}
        Let $x \in \bigcap_{\alpha\in A_1'}\ker{\alpha}$. We want to show that $\alpha x=0$ for all $\alpha\in A_1$. So let $\alpha$ be any element of $A_1$, i.e., $\alpha=\J(ab)-\J(a)\J(b)$ for certain $a,b\in A$. Write $a=a_1+ia_2$ and $b=b_1+ib_2$ with $a_1,a_2,b_1,b_2\in A$. Then
            \begin{align}
                \alpha &= \J\big((a_1+ia_2)(b_1+ib_2)\big)-\J(a_1+ia_2)\J(b_1+ib_2) \nonumber \\
                &= \!\begin{multlined}[t]
                    \big(\J(a_1b_1)-\J(a_1)\J(b_1)\big) + i\big(\J(a_1b_2)-\J(a_1)\J(b_2)\big) \\ + i\big(\J(a_2b_1)-\J(a_2)\J(b_1)\big) - i\big(\J(a_2b_2)-\J(a_2)\J(b_2)\big)
                \end{multlined} \nonumber \\
                &= \alpha_1 + i\alpha_2 + i\alpha_3 - \alpha_4,
            \end{align}
        where
        \begin{alignat}{2}
            \alpha_1&=\big(\J(a_1b_1)-\J(a_1)\J(b_1)\big), &\quad \alpha_2&=\big(\J(a_1b_2)-\J(a_1)\J(b_2)\big), \nonumber \\
            \alpha_3&=\big(\J(a_2b_1)-\J(a_2)\J(b_1)\big), &\quad \alpha_4&=\big(\J(a_2b_2)-\J(a_2)\J(b_2)\big).
        \end{alignat}
        Because $x \in \bigcap_{\alpha\in A_1'}\ker{\alpha}$ and $\alpha_1,\alpha_2,\alpha_3,\alpha_4\in A_1'$ we can now conclude that
        \begin{equation}
            \alpha x = (\alpha_1 + i\alpha_2 + i\alpha_3 -\alpha_4)x = 0.
        \end{equation}
        This shows that $\bigcap_{\alpha\in A_1'}\ker{\alpha} = \bigcap_{\alpha \in A_1}\ker{\alpha}$ and hence $q_1=q_1'$, which proves the claim. 
    
    \noindent Now define $p_\alpha = [\ovl{\ran\, \alpha}]$, i.e., the range projection of $\alpha$, and define the projections
    \begin{align}\label{eqn:q_1,q_2,q_3}
        e_1 &= \inf\{1_H - p_\alpha \mid \alpha \in A_1 \}; \nonumber \\ 
        e_2 &= \inf\{1_H - p_\alpha \mid \alpha \in A_2 \}; \nonumber \\
        e_3 &= \inf\{1_H - p_\alpha \mid \alpha \in A_1\cup A_2 \}.
    \end{align}
    \emph{Claim:}
        We have $q_1=e_1$, $q_2=e_2$ and $q_3=e_3$.
\smallskip

\noindent  
    \emph{Proof of claim:}
        We prove that $q_1=e_1$; the other two identities are proven similarly. We will show that $q_1=e_1$ by proving that $q_1 \leq e_1$ and $e_1 \leq q_1$.
        We first show that $q_1 \leq e_1$. Let 
        \beq
        \alpha=\J(ab)-\J(a)\J(b) \in A_1.\eeq
         We can decompose $H$ as
        \begin{equation}
            H=\ovl{\ran\, {\alpha}}\oplus\ovl{\ran\, {\alpha}}^\perp=\ran\, {p_\alpha}\oplus\big(\ran\, {p_\alpha}\big)^\perp \nn\\
 =\ran\, {p_\alpha}\oplus\ker{p_\alpha^*}=\ran\, {p_\alpha}\oplus\ker{p_\alpha}.
        \end{equation}
        Let $x\in H$ be any element of $H$ and decompose it as $x=y+z$ with $y\in\ran\, {p_\alpha}$ and $z\in\ker{p_\alpha}$. Note that $\alpha^*=\J(b^*a^*)-\J(b^*)\J(a^*)\in A_1$ and
        \begin{equation}
            \ran\, {p_\alpha}\cap\ker{\alpha^*} = \ovl{\ran\, {\alpha}} \, \cap (\ran\, {\alpha})^\perp = (\ran\, {\alpha})^{\perp\perp} \cap (\ran\, {\alpha})^\perp = \{0\}.
        \end{equation}
        So for $0 \neq y\in\ran\, {p_\alpha}$ we have $y\notin\ker{\alpha^*}$ and thus $q_1y=0$. Then
        \begin{align}
            q_1x &= q_1(y+z) = q_1y + q_1z = 0 + q_1z 
            = q_1(y-y) + q_1(z-0) =  q_1(1_H-p_\alpha)y + q_1(1_H-p_\alpha)z \nonumber \\
            &= q_1(1_H-p_\alpha)(y+z) = q_1(1_H-p_\alpha)x.
        \end{align}
        This shows that $q_1\leq(1_H-p_\alpha)$ for all $\alpha\in A_1$, and hence
        \begin{equation}
            q_1 \leq \inf\{1_H - p_\alpha \mid \alpha \in A_1 \}=e_1.
        \end{equation}
        So it remains to be shown that $e_1 \leq q_1$. To this end, note that for all
$ \alpha=\big(\J(ab)-\J(a)\J(b)\big)\in A_1'$,
        \begin{align}
            \alpha^* &= \J(ba) - \J(b)\J(a) = \J(ab+ba)-\J(ab)-\J(b)\J(a) 
            =-\big(\J(ab)-\J(a)\J(b)\big)=-\alpha.
        \end{align}
        Using this relation, we see that
        \begin{equation}\label{eqn:ker=ran_perp}
            \ker{\alpha}=\ker{(-\alpha)}=\ker{\alpha^*}=\big(\ran\, {\alpha}\big)^\perp.
        \end{equation}
      Using the relation above and the previous claim, we now decompose $H$ as
        \begin{align}
            H &= \ran\, \, {q_1} \oplus \big(\ran\, \, {q_1}\big)^\perp
            = \bigcap_{\alpha\in A_1'}\ker{\alpha} \oplus \Big(\bigcap_{\alpha\in A_1'}\ker{\alpha}\Big)^{\perp}
            = \bigcap_{\alpha\in A_1'}\big(\ran\, \, {\alpha}\big)^\perp \oplus \Big(\bigcap_{\alpha\in A_1'}\big(\ran\, \, {\alpha}\big)^{\perp}\Big)^\perp \nonumber \\
            &= \bigcap_{\alpha\in A_1'}\big(\ran\, \, {\alpha}\big)^\perp \oplus \Big(\bigcup_{\alpha \in A_1'}\ran\, \, {\alpha}\Big)^{\perp\perp} = \bigcap_{\alpha\in A_1'}\big(\ran\, \, {\alpha}\big)^\perp \oplus \ovl{\bigcup_{\alpha \in A_1'}\ran\, \, {\alpha}}.
        \end{align}
        Let $x\in H$ be any element of $H$ and decompose it as $x=y+z$ with
        \begin{align}
            y\in \ran\, \, {q_1}=\bigcap_{\alpha\in A_1'}(\ran\, \, {\alpha})^\perp; &&
            z\in \big(\ran\, \, {q_1}\big)^\perp=\ovl{\bigcup_{\alpha \in A_1'}\ran\, \, {\alpha}}.
        \end{align}
        Then it follows immediately that
        \begin{equation}
            q_1 x = q_1(y+z) = y.
        \end{equation}
        Since $z\in \ovl{\bigcup_{\alpha \in A_1'}\ran\, \, {\alpha}}$ there are $z_n\in \bigcup_{\alpha \in A_1'}\ran\, \, {\alpha}$ such that $z_n$ converges to $z$ in $H$. Now let $\alpha_n \in A_1'$ be such that $z_n \in \ran\, \, {\alpha_n}$. Because $e_1 \leq 1_H-p_\alpha$ for all $\alpha \in A_1'$, we have $e_1=e_1(1-p_\alpha)$ for all $\alpha\in A_1'$, and hence:
        \begin{align}
            e_1 x &= e_1(y+z) = e_1y +\lim_{n\to\infty}e_1z_n = e_1y + \lim_{n\to\infty}e_1(1_H-p_{\alpha_n})z_n \nonumber \\
            &= e_1y + \lim_{n\to\infty}e_1(z_n-z_n) = e_1y = e_1q_1 x.
        \end{align}
        This shows that $e_1q_1=e_1$ and hence $e_1 \leq q_1$. We conclude that $e_1=q_1$, which proves the claim.
   \smallskip
    
\noindent    \emph{Claim:}
        The projections $q_1$, $q_2$ and $q_3$ are central in the von Neumann algebra $A''$.
\smallskip

\noindent    \emph{Proof of claim:}
        We prove the statement for $q_1$; the other statements are proven analogously. First, $q_1=\inf\{1_H-p_\alpha \mid \alpha\in A_1\}$,
    where  $p_\alpha$ is the range projection of $\alpha\in A_1 \subseteq  A$, hence $p_\alpha\in  A''$, and therefore also
 $q_1\in A''$.       
        Second, we show that $q_1\in A'$; the proofs for $q_2$ and $q_3$ are analogous. We want to show that $q_1 A= Aq_1$ for all $a\in A$. Decompose $H$ as follows:
        \begin{equation}
            H = \bigcap_{\alpha\in A_1}\ker{\alpha} \oplus \Big(\bigcap_{\alpha\in A_1}\ker{\alpha}\Big)^\perp.
        \end{equation}
        Note that the range space of $q_1$ is invariant under $ A$, i.e.,
        \begin{equation}
            \J(a)\bigcap_{\alpha \in A_1}\ker{\alpha}\subseteq\bigcap_{\alpha \in A_1}\ker{\alpha},
        \end{equation}
        for all $a\in A$. Indeed, let $a\in A$, $x\in \bigcap_{\alpha \in A_1}\ker{\alpha}$, and $\alpha=\J(bc)-\J(b)\J(c)\in A_1$. Then
        \begin{align}
            \alpha\J(a) &= \big(\J(bc)-\J(b)\J(c)\big)\J(a) \nonumber \\
            &= -\big(\J(bca)-\J(bc)\J(a)\big) + \big(\J(bca)-\J(b)\J(ca)\big) + \J(b)\big(\J(ca)-\J(c)\J(a)\big) \nonumber \\
            &= -\alpha_1 + \alpha_2 +\J(b)\alpha_3,
        \end{align}
        where
        \begin{align}
            \alpha_1=\J(bca)-\J(bc)\J(a); &&
            \alpha_2=\J(bca)-\J(b)\J(ca); &&
            \alpha_3=\J(ca)-\J(c)\J(a),
        \end{align}
        and $\alpha_1,\alpha_2,\alpha_3\in A_1$. Because $x\in \bigcap_{\alpha \in A_1}\ker{\alpha}$, it follows immediately that
        \begin{equation}
            \alpha\J(a)x=(-\alpha_1+\alpha_2+\J(b)\alpha_3)x=0,
        \end{equation}
        so $\J(a)x\in\ker{\alpha}$, and hence
        \begin{equation}
            \J(a)x\in \bigcap_{\alpha\in A_1}\ker{\alpha}.
        \end{equation}
        Because $\bigcap_{\alpha \in A_1}\ker{\alpha}$ is invariant under $ A$ we also know that $\Big(\bigcap_{\alpha \in A_1}\ker{\alpha}\Big)^\perp$ is invariant under $ A$. Now let $x\in H$ and write it as $x=y+z$ with $y\in \bigcap_{\alpha\in A_1}\ker{\alpha}$ and $z\in \Big(\bigcap_{\alpha\in A_1}\ker{\alpha}\Big)^\perp$. Then
        \begin{align}
            q_1\J(a)x &= q_1\J(a)(y+z)= \J(a)y + 0 = \J(a)q_1y + \J(a)q_1z = \J(a)q_1x,
        \end{align}
        which shows that $q_1\in A'$. So now we know that $q_1\in  A'' \cap  A'$, which proves that $q_1$ is central in the von Neumann algebra generated by $ A$. This proves the claim.
  \smallskip
    
    \noindent The second claim showed that 
    \begin{align}
        q_1 &= \inf\{1_H - p_\alpha \mid \alpha \in A_1 \}, \nonumber \\
        q_2 &= \inf\{1_H - p_\alpha \mid \alpha \in A_2 \}, \nonumber \\
        q_3 &= \inf\{1_H - p_\alpha \mid \alpha \in A_1\cup A_2 \},
    \end{align}
    which in turn means that
    \begin{align}\label{eqn:p_1,p_2,1-p_3}
        1_H-q_1 &= \sup\{p_\alpha \mid \alpha \in A_1 \}, \nonumber \\
        1_H-q_2 &= \sup\{p_\alpha \mid \alpha \in A_2 \}, \nonumber \\
        1_H-q_3 &= \sup\{p_\alpha \mid \alpha \in A_1\cup A_2 \}.
    \end{align}
    Now define
    \begin{equation}
        p_1 = 1_H-q_2, \quad p_2 = 1_H-q_1, \quad p_3 = q_1.
    \end{equation}
    We now show that $p_1$, $p_2$ and $p_3$ satisfy the requirements of the theorem. The third claim showed that $p_1$, $p_2$ and $p_3$ are central projections. We now show that they are mutually orthogonal and that $p_1+p_2+p_3=1_H$. Let $\alpha\in A_1$ and $\beta\in A_2$. By Lemma \ref{lemma:A_1*A_2=0} we have $\alpha^*\beta=0$, so 
    \beq
    \ran\, \, {\beta} \subseteq \ker{\alpha^*}=(\ran\, \, {\alpha})^\perp,
    \eeq
     and hence
    \begin{equation}
        \ovl{\ran\, {\beta}} \subseteq \ovl{(\ran\, {\alpha})^\perp}.
    \end{equation}
    Because $(\ran\, {\alpha})^\perp$ is closed we have $\ovl{(\ran\, {\alpha})^\perp}=(\ran\, {\alpha})^\perp$, and
    \begin{equation}
        (\ran\, {\alpha})^\perp=\big((\ran\, {\alpha})^{\perp\perp}\big)^\perp=\ovl{\ran\, {\alpha}}^\perp,
    \end{equation}
    which implies $p_\alpha p_\beta=0$ for all $\alpha\in A_1$ and $\beta \in A_2$. Because $\{p_\alpha \mid \alpha \in A_1 \}$ and $\{p_\beta \mid \beta \in A_2 \}$ are bounded sets in $B(H)$, where multiplication is $\sg$-strongly \emph{jointly} continuous, 
    it follows that
    \begin{equation}\label{eqn:p_1p_2=0}
        p_2p_1 = \big(\sup_{\alpha\in A_1}p_\alpha\big)\big(\sup_{\beta\in A_2}p_\beta\big) = \sup_{\alpha\in A_1}\sup_{\beta\in A_2}\big(p_\alpha p_\beta\big) = 0.
    \end{equation}
    This shows that $p_1$ and $p_2$ are mutually orthogonal and hence that $p_1+p_2$ is a projection. Because $A_1\subseteq A_1\cup A_2$ and $A_2 \subseteq A_1\cup A_2$, we can immediately conclude from \eqref{eqn:p_1,p_2,1-p_3} that
    \begin{equation}\label{eqn:projection_inequalities}
        p_1\leq 1_H-p_3 \quad \text{and} \quad p_2\leq 1_H-p_3.
    \end{equation}
    This shows that
    \begin{equation}
        (p_1+p_2)(1_H-p_3) = p_1(1_H-p_3)+p_2(1_H-p_3) = p_1+p_2,
    \end{equation}
    which implies
    \begin{equation}
        p_1+p_2 \leq 1_H-p_3.
    \end{equation}
  As in \er{eqn:p_1p_2=0}, it follows that for every $\alpha \in A_1$ and $\beta \in A_2$
    \begin{equation}\label{eqn:p_1alpha=0_and_p_2beta=0}
        p_1\alpha = 0, \quad p_2\beta=0,
    \end{equation}
    and hence
    \begin{equation}\label{eqn:p_1p_alpha=0_and_p_2p_beta=0}
        p_1p_\alpha = 0, \quad p_2p_\beta=0.
    \end{equation}
    Because $p_\alpha \leq p_2$ for all $\alpha\in A_1$ and $p_\alpha \leq p_1$ for all $\alpha \in A_2$, the above relation shows that
    \begin{equation}
        \big(p_1+p_2\big)p_\alpha=p_\alpha,
    \end{equation}
    for all $\alpha \in A_1 \cup A_2$. Hence
    \begin{equation}
        p_\alpha \leq p_1+p_2,
    \end{equation}
    which shows that
    \begin{equation}
        1_H-p_3=\sup_{\alpha \in A_1 \cup A_2}p_\alpha \leq p_1+p_2.
    \end{equation}
    Because $p_1+p_2 \leq 1_H-p_3$ and $1_H-p_3 \leq p_1+p_2$ we have
    \begin{equation}
        p_1+p_2 = 1_H-p_3,
    \end{equation}
    and hence
    \begin{equation}
        p_1+p_2+p_3 = 1_H.
    \end{equation}
    Now use this relation to see that
    \begin{equation}
        p_1p_3=p_1(1_H-p_1-p_2)=p_1-p_1-p_1p_2=0,
    \end{equation}
    and similarly $p_2p_3=0$. So the projections $p_1$, $p_2$ and $p_3$ are mutually orthogonal. It is clear from  \eqref{eqn:p_1,p_2,1-p_3} and  \eqref{eqn:p_1alpha=0_and_p_2beta=0} that for all $a,b\in A$,
    \begin{equation}
        \J(ab)p_1-\J(a)p_1\J(b)p_1 = (\J(ab)-\J(a)\J(b))p_1=0,
    \end{equation}
    which shows that $\J(\cdot)p_1$ is a homomorphism. Now suppose that $p_1 \neq 0$. By equation \eqref{eqn:p_1,p_2,1-p_3} this means that there exists an $\alpha=\J(ab)-\J(b)\J(a)\in A_2$ such that $\alpha \neq 0$. Then
    \begin{equation}
        \J(ab)p_1-\J(b)p_1\J(a)p_1 = \alpha p_1 = p_1p_\alpha \alpha = p_\alpha \alpha = \alpha \neq 0,
    \end{equation}
    which shows that $\J(\cdot)p_1$ is not an anti-homomorphism. The proof that $\J(\cdot)p_2$ is an anti-homomor\-phism and not a homomorphism and the proof that $\J(\cdot)p_3$ is a homomorphism as well as an anti-homomorphism are similar. Lastly, suppose that there is a central projection $p$ such that $\J(\cdot)p$ is a homomorphism as well as an anti-homomorphism. Then
    \begin{equation}
        p(\J(ab)-\J(a)\J(b))=0=p(\J(ab)-\J(b)\J(a)),
    \end{equation}
    for all $a,b \in A$. This shows that $\ran\, {p}\subseteq \ker{\alpha}$ for all $\alpha\in A_1\cup A_2$ and hence:
    \begin{equation}
        \ran\, {p}\subseteq \bigcap_{\alpha\in A_1\cup A_2}\ker{\alpha}=\ran\, {p_3}.
    \end{equation}
    This implies that $p \leq p_3$, which proves that $p_3$ is the largest central projection such that $\J(\cdot)p_3$ is a homomorphism, as well as an anti-homomorphism.
    
    Finally, we show that the conditions in the theorem determine $p_1$, $p_2$ and $p_3$ uniquely. So assume that $p_1$, $p_2$ and $p_3$ satisfy the conditions of the theorem and let $q_1$, $q_2$ and $q_3$ be as in  \eqref{eqn:p_1,p_2,1-p_3}. We show that $p_1=1_H-q_2$, $p_2=1_H-q_1$ and $p_3=q_3$. By the previous argument, we have $p_3 \leq q_3$. Because $p_3$ is the largest central projection such that $\J(\cdot)p_3$  is a homomorphism, as well as an anti-homomorphism, we also have $q_3 \leq p_3$. This proves that $p_3=q_3$. Furthermore, $p_1+p_2+p_3=1_H$, which implies
    \begin{align}
        p_1 &\leq 1_H-p_3; \\
        p_2 & \leq 1_H-p_3; \\
        p_1+p_2 &=1_H-p_3.
    \end{align}
    Let $a,b\in A$. Because $p_1$ is a central projection such that $\J(\cdot)p_1$ is a homomorphism, we have
    \begin{equation}
        \J(ab)p_1-\J(a)p_1\J(b)p_1 = (\J(ab)-\J(a)\J(b))p_1=0,
    \end{equation}
    and hence
    \begin{equation}
        p_1p_\alpha = 0,
    \end{equation}
    where $\alpha = \J(ab)-\J(a)\J(b) \in A_1$. This shows that 
    \begin{equation}
        p_1(1_H-p_\alpha) = p_1,
    \end{equation}
    for all $\alpha \in A_2$, and therefore,
    \begin{equation}
        p_1 \leq \inf\{1_H-p_\alpha \mid \alpha \in A_1 \} = q_1.
    \end{equation}
    Similarly, $p_2 \leq q_2$. This in turn implies that $1_H-q_1 \leq 1_H-p_1$ and $1_H-q_2 \leq 1_H-p_2$. Therefore,
    \begin{equation}
        1_H-q_2 = (1_H-q_2)(1_H-p_3) \leq (1_H-p_2)(1_H-p_3) = 1_H-p_2-p_3 = p_1,
    \end{equation}
    and similarly,
    \begin{equation}
        1_H-q_1 \leq p_2,
    \end{equation}
    where we used the fact that $(1_H-q_2)p_3=(1_H-q_1)p_3=0$. When we combine these inequalities we find that
    \begin{equation}
        1_H-p_3 = (1_H-q_1)+(1_H-q_2) \leq p_1 + p_2 = 1_H-p_3,
    \end{equation}
    so the equalities must hold, i.e.,
    \begin{align}
        1_H-q_2 &= p_1, \\
        1_H-q_1 &=p_2.
    \end{align}
This also shows uniqueness of $p-1$ and $p_2$ and the proof of Lemma \ref{Thomsen} is finished \hfill $\Box$
\bigskip
 
\noindent \textbf{Acknowledgement.} The authors are indebted to C. Akemann and F.W. Shultz for  help with some intricate mathematical points, as well as to M.\ Frank for some references they had missed. 
\newpage
\addcontentsline{toc}{section}{References}
\begin{small}

\end{small}
\end{document}

%% file: Bohrificationpreamble.tex
\newcommand{\id}[1]{\ensuremath{\mathrm{id}}}

\newcommand{\quar}{\mbox{\footnotesize $\frac{1}{4}$}}
\newcommand{\half}{\mbox{\footnotesize $\frac{1}{2}$}}
\newcommand{\hi}[1]{\emph{\textbf{#1}}}
\newcommand{\qm}{quantum mechanics}
\newcommand{\er}{\eqref}
 
\newcommand{\beq}{\begin{equation}}
\newcommand{\eeq}{\end{equation}} 
\newcommand{\bea}{\begin{eqnarray}}
\newcommand{\eea}{\end{eqnarray}} \newcommand{\nn}{\nonumber}

\newcommand{\ovl}{\overline}

 \newcommand{\til}{\tilde}
\newcommand{\raw}{\rightarrow}

\newcommand{\sa}{\ensuremath{_{\mathrm{sa}}}}

\newcommand{\lraw}{\leftrightarrow}

\newcommand{\ot}{\otimes} 
\newcommand{\la}{\langle} \newcommand{\ra}{\rangle}
\newcommand{\ran}{\mathrm{ran}}
 
\newcommand{\x}{\times}

\newcommand{\tp}{transition probability}
\newcommand{\tpies}{transition probabilities}

\newcommand{\ca}{C*-algebra} 
 \newcommand{\rep}{representation}

\newcommand{\Hs}{Hilbert space}

\newcommand{\sta}{$\mbox{}^*$-algebra}

   \newcommand{\vna}{von
Neumann algebra}

\newcommand{\al}{\alpha} 
\newcommand{\gm}{\gamma} 
\newcommand{\dl}{\delta} \newcommand{\Dl}{\Delta}
 \newcommand{\varep}{\varepsilon}

\newcommand{\lm}{\lambda} 
\newcommand{\rh}{\rho} \newcommand{\sg}{\sigma}
 \newcommand{\ta}{\tau} 
 \newcommand{\phv}{\varphi}
\newcommand{\ch}{\ch} \newcommand{\ps}{\psi} 
\newcommand{\om}{\omega} \newcommand{\Om}{\Omega}

\newcommand{\inv}{^{-1}}

\newcommand{\Tr}{\mbox{\rm Tr}\,}

 \newcommand{\K}{\mathbb{K}}

\newcommand{\CC}{{\mathcal C}} 
\newcommand{\CE}{{\mathcal E}}

 \newcommand{\CS}{{\mathcal S}}
 \newcommand{\CP}{{\mathcal P}}

\newcommand{\C}{{\mathbb C}} 
 \newcommand{\R}{{\mathbb R}}
\newcommand{\T}{{\mathbb T}} \newcommand{\Z}{{\mathbb Z}}
  \makeatletter
 
\makeatletter
\def\moverlay{\mathpalette\mov@rlay}
\def\mov@rlay#1#2{\leavevmode\vtop{%
   \baselineskip\z@skip \lineskiplimit-\maxdimen
   \ialign{\hfil$\m@th#1##$\hfil\cr#2\crcr}}}
\newcommand{\charfusion}[3][\mathord]{
    #1{\ifx#1\mathop\vphantom{#2}\fi
        \mathpalette\mov@rlay{#2\cr#3}
      }
    \ifx#1\mathop\expandafter\displaylimits\fi}
\makeatother

%% file: Wignerv2.bbl
\begin{thebibliography}{99}
 \bibitem{} Alfsen, E.M., Hanche-Olsen, H., Shultz, F.W. (1980).
 State spaces of C*-algebras. \emph{Acta Mathematica} 144, 267--305. 
\bibitem{} Alfsen, E.M.,  Shultz, F.W. (2001). \emph{State Spaces of Operator Algebras}. Basel: Birkh\"{a}user.
\bibitem{}Alfsen, E.M.,  Shultz, F.W. (2003).  \emph{Geometry of State Spaces of Operator Algebras}. Basel: Birkh\"{a}user.      
 \bibitem{} Bargmann, V. (1964). Note on Wigner's Theorem on symmetry operations. \emph{Journal of Mathematical Physics} 5, 862--868.   
     \bibitem{}   Baki\'{c}, D., Gulja\v{s}, B. (2003). Wigner's theorem in a class of Hilbert C*--modules.
     \emph{Journal of Mathematical Physics} 44, 2186--2191. 
 \bibitem{}Birkhoff, G. \&  Kreyszig, E. (1984). The establishment of functional analysis. \emph{Historia Mathematica} 11, 258--321.
\bibitem{}Bonolis, L. (2004). From the rise of the group concept to the stormy onset of group theory in the new
quantum mechanics: A saga of the invariant characterization of physical objects, events and theories.
\emph{Nuovo Cimento Rivista}  27. DOI:
\texttt{10.1393/ncr/i2004-10006-4}.
\bibitem{} Bourbaki, N. (1989). \emph{General Topology: Chapters 1--4}. Berlin: Springer.
\bibitem{}Bratteli, O.,   Robinson, D.W. (1987). \emph{Operator
Algebras and Quantum Statistical Mechanics. Vol.\ I: C*- and
W*-Algebras, Symmetry Groups, Decomposition of States}. 2nd
Ed. Berlin: Springer.
   \bibitem{}Cassinelli, G., De Vito, E., Lahti, P.J., Levrero, A. (1997).
   Symmetry groups in \qm\ and the theorem of Wigner on the symmetry transformations.
   \emph{Reviews in Mathematical Physics} 9, 921--941. 
   \bibitem{}Cassinelli, G., De Vito, E., Lahti, P.J., Levrero, A. (2004). \emph{The Theory of Symmetry Actions in Quantum Mechanics}. \emph{Lecture Notes in Physics} 654. Berlin: Springer-Verlag.
      \bibitem{} Chevalier, G. (2007). Wigner's Theorem and its generalizations. \emph{Handbook of Quantum Logic and Quantum Structures: Quantum Structures}, pp. 429--475. Engesser, K., Gabbay, D.M., Lehmann, D., eds. Amsterdam: Elsevier. 
   \bibitem{} Connes, A. (1974). 
Caract\'{e}risation des espaces vectoriels ordonn\'{e}s sous-jacents aux alg\`{e}bres de von Neumann.
\emph{Annales de l'institut Fourier} 24, 121--155.
   \bibitem{} Connes, A. (1975). Classification of automorphisms of hyperfinite factors of type  {\sc ii}$\mbox{}_1$ and  {\sc ii}$\mbox{}_{\infty}$ and application to type {\sc iii} factors. \emph{ Bulletin of the American Mathematical Society} 81, 1090--1092. 
    \bibitem{} Dieks, D. (2016). Von Neumann's impossibility proof: Mathematics
in the service of rhetorics. \emph{Studies in History and Philosophy of Modern Physics} 60, 136--148. 
\bibitem{} Dixmier, J. (1977). \emph{C*-algebras}.  Amsterdam: North-Holland.
\bibitem{} Dixmier, J. (1981). \emph{Von Neumann Algebras}.  Amsterdam: North-Holland.
\bibitem{} Duncan, A., Janssen, M. (2013). (Never) Mind your $p$'s and $q$'s: Von Neumann versus Jordan on the foundations of quantum theory. \emph{The European Physical Journal H} 38, 175--259.
\bibitem{}Freed, D.S. (2012). On Wigner's Theorem. 
\emph{Geometry \& Topology Monographs} 18, 83--89. 
 \bibitem{} Hanche-Olsen, H.,  St\o rmer, E. (1984). \emph{Jordan Operator Algebras}. Boston: Pitman. 
\bibitem{} Herstein, I.N. (1956). Jordan homomorphisms. \emph{Transactions of the American Mathematical Society} 81, 331--341.
 \bibitem{} Hunziker, W. (1972). A note on symmetry operations in quantum mechanics.
 \emph{Helvetica Physica Acta} 45, 233--236. 
\bibitem{} Jacobson, N., Rickart, C.E. (1950). Jordan homomorphisms of rings. \emph{Transactions of the American Mathematical Society} 69, 479--502.
\bibitem{}  Jordan, P. (1933) \"{U}ber Verallgemeinerungsm\"{o}glichkeiten des Formalismus der Quantenmechanik.
\emph{Nachrichten von der Gesellschaft der Wissenschaften zu G\"{o}ttingen, Mathematisch-Physikalische Klasse}
 41, 209--217.
 \bibitem{} Jordan, P., von Neumann, J., Wigner, E.P. (1934). 
On an algebraic generalization of the quantum mechanical formalism. 
\emph{Annals of Mathematics 35, 29--64.}
\bibitem{}
Kadison, R.V. (1951). Isometries of operator algebras.  \emph{Annals of Mathematics} 54, 325--338.
\bibitem{}
Kadison, R.V. (1957). Unitary invariants for representations of operator algebras. \emph{Annals of Mathematics} 66, 304--379. 
\bibitem{}
Kadison, R.V. (1965). Transformation of states in operator theory and dynamics. \emph{Topology} 3, 177-198.
\bibitem{}Kadison, R.V.,  Ringrose, J.R. (1983). \emph{Fundamentals of the Theory of Operator Algebras. Vol. 1: Elementary Theory}.   New York: Academic Press.
 \bibitem{} Keller, K.J., Papadopoulos, N.A.,  Reyes-Lega, A.F. (2008).
 On the realization of symmetries in quantum mechanics. \emph{Mathematische Semesterberichte}  55, 149--160.
 \bibitem{}  Landsman, K. (1998). \emph{Mathematical Topics Between Classical and Quantum Mechanics}. New York: Springer. 
 \bibitem{}  Landsman, K. (2017). \emph{Foundations of Quantum Theory: From Classical Concepts to Operator Algebras}.
 Cham: Springer Open.  \verb#https://www.springer.com/gp/book/9783319517766#.
  \bibitem{}  Landsman, K. (2019). Quantum theory and functional analysis. \emph{Oxford Handbook of the History of Interpretations and Foundations of Quantum Mechanics}, ed.\ O. Freire. Oxford: Oxford University Press, to appear in 2021. 
  \verb#https://www.math.ru.nl/~landsman/QTFA.pdf#.
  \bibitem{}Mielnik, B. (1968). Geometry of quantum states. 
\emph{Communications in Mathematical Physics}  9, 55--80.
 \bibitem{}  Moln\'{a}r, L. (1998). 
An algebraic approach to Wigner's unitary-antiunitary theorem. \emph{Journal of the Australian Mathematical Society}
65, 354--369. 
 \bibitem{}  Moln\'{a}r, L. (1999). A generalization of Wigner's unitary-antiunitary theorem to Hilbert modules.
 \emph{Journal of Mathematical Physics} 40, 5544--5554. 
 \bibitem{}  Moln\'{a}r, L. (2000).  Wigner-type theorem on symmetry transformations in type II factors.
\emph{International Journal of Theoretical Physics} 1463--1466.
 \bibitem{}Moretti, V. (2013). \emph{Spectral Theory and Quantum Mechanics}. Mailand: Springer.
%  \bibitem{} Mori. M. (2018). Isometries between projection lattices of von Neumann algebras.
 % \emph{Journal of Functional Analysis} 176, 3511--3528. 
 \bibitem{}Neumann, J. von (1932). \emph{Mathematische Grundlagen der Quantenmechanik.} Berlin: Springer.
 \bibitem{}Neumann, J.  von (1981).  
Continuous geometries with a transition probability. Edited by Halperin, I.S.
\textit{Memoirs of the American 
Mathematical Society}  252, 1--210.
\bibitem{}Neumann, J. von, Wigner, E.P. (1928). Zur Erkl\"{a}rung einiger Eigenschaften der Spektren aus der Quantenmechanik
des Drehelektrons. \emph{Zeitschrift f\"{u}r Physik}  A49, 73--94.
\bibitem{}Pedersen, G.K. (1979). \emph{\ca s and their Automorphism Groups}. London: Academic Press.
\bibitem{} Rang, K. (2019).  \emph{Symmetries in algebraic quantum theory}. M.Sc Thesis, Radboud University.
\verb#https://www.math.ru.nl/~landsman/Kitty2019.pdf#.
\bibitem{}R\'{e}dei, M. (1996). Why John von Neumann did not like the \Hs\ formalism of \qm\ (and what he liked instead).
\emph{Studies in History and Philosophy of Modern Physics} 27, 493--510. 
\bibitem{}  R\'{e}dei, M., St\"{o}ltzner, eds.\ (2001).   \emph{John von Neumann and the Foundations of Quantum Physics}.
   Dordrecht: Kluwer. 
   \bibitem{}
Roberts, J.E. and G. Roepstorff (1969).  Some basic concepts of algebraic quantum theory. \emph{Communications in Mathematical Physics} 11, 321--338.
\bibitem{}Scholz, E. (2006). Introducing groups into quantum theory (1926--1930). \emph{Historia Mathematica} 33, 440--490. 
  \bibitem{}Shultz, F.W. (1982). Pure states as dual objects for \ca s. \emph{Communications in Mathematical Physics}  82, 497--509.
        \bibitem{} 
    Simon, B. (1976). Quantum dynamics: from automorphism to Hamiltonian.  \emph{Studies
in Mathematical Physics: Essays in Honor of Valentine Bargmann}, pp.\ 327--349. Lieb, E., Simon, B., Wightman, A.S., eds.
 Princeton: Princeton University Press. 
         \bibitem{} 
    Simon, R., Mukunda, N., Chaturvedi, S., Srinivas, V. (2008).
    Two elementary proofs of the Wigner theorem on symmetry in quantum mechanics. \emph{Physics Letters A} 372, 6847--6852.
          \bibitem{}  Takesaki, M. (2002). \emph{Theory of Operator Algebras. Vol.\ I}.  New York: Springer-Verlag.
 \bibitem{} Thomsen, K. (1982). Jordan-morphisms in \sta s. \emph{Proceedings of the American Mathematical Society}
86, 283--286. 
\bibitem{} Uhlhorn, U. (1962). Representation of symmetry transformations in quantum mechanics. \emph{Arkiv f\"{o}r Fysik} 23, 307--340. 
\bibitem{}Wigner, E.P. (1931). \emph{Gruppentheorie und ihre Anwendung auf die Quantenmechanik der Atomspektren}.
Wiesbaden: Vieweg. 
\end{thebibliography}
